\documentclass[twocolumn,pre,groupedaddress,showpacs,superscriptaddress,amsmath]{revtex4-1}
\usepackage{subfigure}

\usepackage{epsfig,amsfonts,amsbsy,graphicx,amsfonts}

\usepackage{color}

\graphicspath{{./figure/}}
\setcitestyle{super}
   
\begin{document}

\title{Emergence of collective oscillations in adaptive cells}
 
\author{Shou-Wen Wang}
\email{Correspondence: shouwen$\_$wang@hms.harvard.edu}
\affiliation{Beijing Computational Science Research Center, Beijing, 100094, China}
\affiliation{Department of Engineering Physics, Tsinghua University, Beijing, 100086, China}
\affiliation{Department of Systems Biology, Harvard Medical School, Boston, MA 02115, USA}

\author{Lei-Han Tang}
\email{Correspondence: lhtang@csrc.ac.cn}
\affiliation{Beijing Computational Science Research Center, Beijing, 100094, China}
\affiliation{Department of Physics and Institute of Computational and Theoretical Studies, Hong Kong Baptist University, Hong Kong, China}
\affiliation{State Key Laboratory of Environmental and Biological Analysis, Hong Kong Baptist University, Hong Kong, China}

\date{\today}

\begin{abstract}

{\color{black} Collective oscillation of cells in a population has been reported under diverse biological contexts and with vastly
different molecular constructs.  Could there be common principles similar to those that govern spontaneous oscillation 
in mechanical or electrical systems? Here, we answer this question in the affirmative by categorising the response 
of individual cells against a time-varying signal.  A positive intracellular signal relay of sufficient gain from participating cells
is required to sustain the oscillations, together with phase matching. The two conditions yield quantitative predictions 
for the onset cell density and frequency in terms of measured single-cell and signal response functions. 
Through mathematical constructions, we show that cells that adapt to a constant stimulus fulfil the phase requirement
by developing a leading phase in an ``active'' frequency window that enables cell-to-signal energy flow. 
Analysis of dynamical quorum sensing in several cellular systems with increasing biological complexity 
reaffirms the pivotal role of adaptation in powering oscillations in an otherwise dissipative cell-to-cell communication channel. 
The physical conditions identified can be used to design synthetic oscillatory systems.}

\end{abstract}

\maketitle

{\color{black} Homogeneous cell populations are able to exhibit a rich variety of organised behaviour, among them periodic oscillations. }
During mound formation of starved social amoebae, cyclic AMP waves guide migrating cells towards the high density
region~\cite{schaap2011evolutionary,gregor2010onset,sgro2015intracellular,kamino2017fold,bretschneider2016progress}.
Elongation of the vertebrate body axis proceeds with a segmentation clock~\cite{hubaud2014signalling,hubaud2017excitable}.  Multicellular pulsation has also been observed in nerve tissues~\cite{buzsaki2006rhythms}, 
during dorsal closure in late stage drosophila embryogenesis~\cite{Solon2009DorsalClosure,sokolow2012cell,dierkes2014spontaneous,julicher1997spontaneous},  
and more~\cite{ko2010emergence}. 
In these  examples, communication through chemical or mechanical signals is essential to activate quiescent cells.
Dubbed ``dynamical quorum sensing'' (DQS) to emphasise the role of increased cell
density in triggering the auto-induced oscillations, this class of behaviour lies outside the well-known 
Kuramoto paradigm of oscillator synchronisation~\cite{kuramoto2012chemical,strogatz2004sync}. 

Interestingly, auto-induced oscillations have also been reported in situations without an apparent biological function. 
A case in point is otoacoustic emission (OAE), where a healthy human ear emits sound spontaneously in a silent environment~\cite{gold1948hearing,kemp1978stimulated}.   
Anatomically, sound is generated by hair bundles, the sensory units of hair cells that detect sound with 
ultra-high sensitivity~\cite{Martin1999hairResponse,martin2001comparison,hudspeth2014integrating}.    
Another example is glycolytic oscillations of yeast cells which can be induced across different laboratory conditions~\cite{goldbeter1997biochemical,richard2003rhythm,de2007dynamical,chandra2011glycolytic,gustavsson2012sustained,amemiya2015collective}.  {\color{black} This type of phenotypic behaviour may not confer benefits to the organism, so their existence is puzzling.}
 
Here, we consider a population of cells ``attempting'' to modulate temporal variations of the extracellular concentration of a 
protein or analyte, or a physical property of their environment, by responding to it.  
The response of a cell to the external property, or ``signal'',  can be mediated by an arbitrary intracellular biochemical network. 
By focusing on the frequency-resolved cellular response, we report a generic condition for collective oscillations 
to emerge, and show that it is satisfied when cells affect the signal in a way that \emph{adapts} to slow environmental variations, i.e., cells respond to signal variation rather than to its absolute level.  
In particular, we prove the existence of an ``active'' frequency regime, where adaptive cells anticipate signal variation and
attempt to amplify the signal. Sustained collective oscillations emerge when a cell population, beyond a critical density, 
communicates spontaneously through such a channel.

We provide a physical explanation of oscillations in terms of energy driven processes, with adaptive cells outputting energy in the active frequency regime upon stimulation.  For mechanical signals, the energy output is directly observable as work on the environment. For chemical signals,  chemical free energy is transferred during the release of molecules into the extracellular medium. 
Together with the measurable response of individual cells, quantitative predictions of the oscillation 
frequency and its dependence on cell density become possible.

The adaptive cellular response highlighted in this work is shown to underlie several known examples of DQS, 
and possibly glycolytic oscillations in yeast cell suspensions. The ubiquity of adaptation~\cite{goentoro2009evidence,
cohen2009dynamics,alon1999robustness,tu2013quantitative,reisert2001response,hohmann2002osmotic,hazelbauer2008bacterial,menini1999calcium,nakatani1991light,mettetal2008frequency,hoeller2014understand}
in biology may also explain the emergence of inadvertent oscillations.  We discuss implications and predictions of this general mechanism at the end of the paper, {\color{black} in connection with previous experimental and modelling work}.\\

\begin{figure}
\centering
\includegraphics[width=8.5cm]{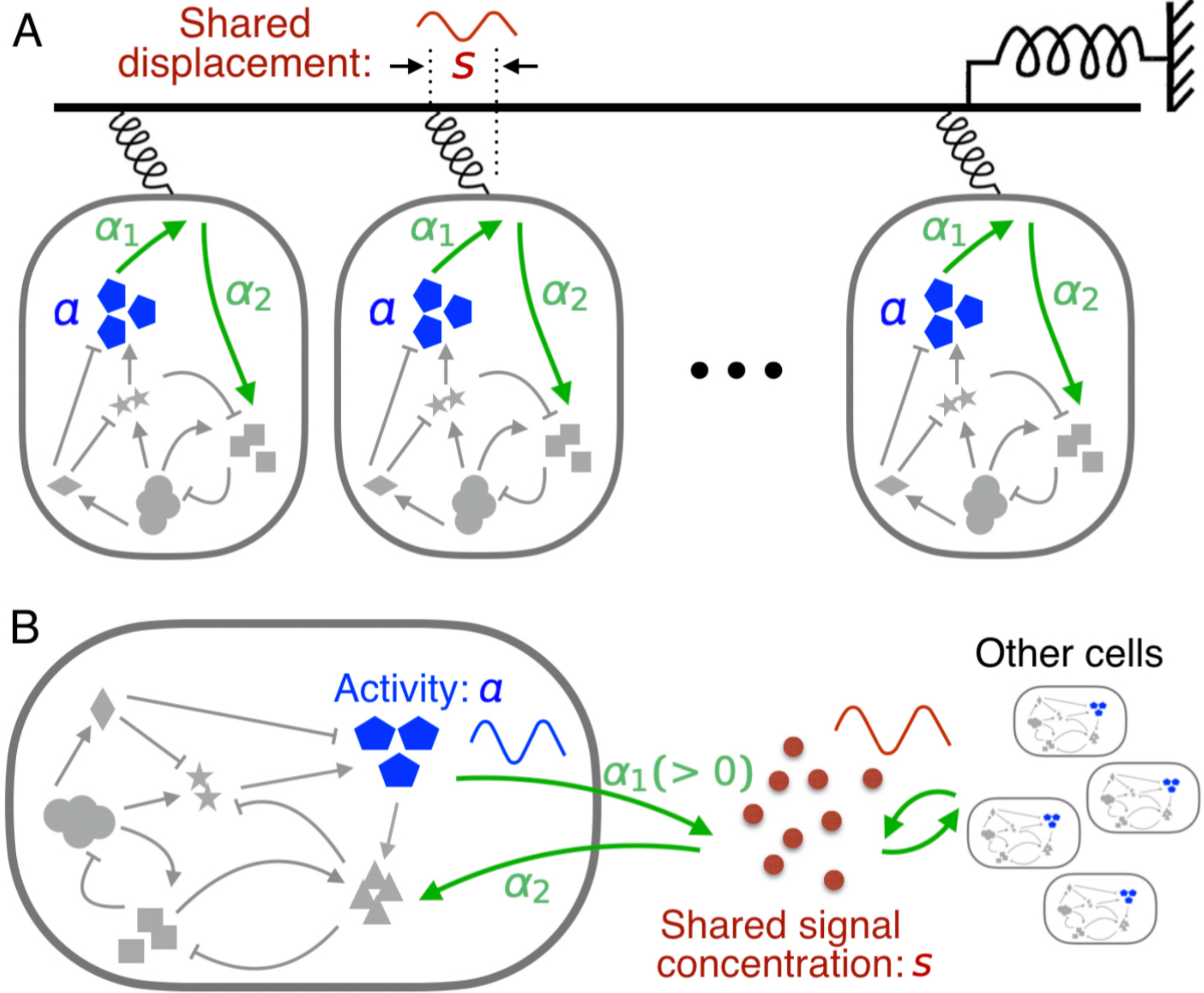}
\caption{Spontaneous oscillations in a communicating cell population.
({\it A})  The scenario of mechanical oscillations where cells communicate via a shared displacement $s$ of the physical environment.  Activity $a$ of a cell against the displacement
is regulated by a hidden intracellular network which responds to $s$ through a mechanical sensor.
({\it B}) Illustration of chemical oscillations where cells interact via a shared extracellular signal $s$.  
The signal is sensed and secreted by individual cells. 
}
\label{fig:scenarios}
\end{figure}

\noindent
{\large \textbf{Results}}\\
\textbf{Necessary conditions for auto-induced collective oscillations.} We begin by considering a scenario of mechanical oscillations, as  illustrated in Fig.~\ref{fig:scenarios}{\it A}. Later we will show that the same results hold for chemical oscillations. The cells are spatially close enough so that they could be regarded as under the same environment. 
Here, the extracellular signal $s$ is taken to be the deformation of the mechanical environment, which is both sensed and modified by participating cells. The ``cell activity'' that affects the environment is 
denoted by a variable $a$. Its dynamics is controlled by an unspecified intracellular regulatory network that responds to $s$ through a mechanical sensor. To see how the intracellular activities might disrupt stasis in an equilibrium state of $s$,
we consider the following Langevin equation: 
 \begin{equation}
 \gamma \dot{s} =F(s)+\sum_{j=1}^N \alpha_1 a_j+\xi.
\label{eq:general-adaptation-4}
 \end{equation}
Here $\gamma$ is the friction coefficient, $F(s)$  the external force that tries to restore the physical environment, $\xi$ the thermal noise, and  the sum represents the total force created by $N$ active cells in a unit volume, whose strength is set by $\alpha_1>0$.  
In general, the cell activity depends on the past history of the signal.  Upon a small change of $s$, the average response of 
the activity of $j$th cell satisfies
\begin{equation}
\langle a_j(t)\rangle=\langle a_j\rangle_{u}+\int_{-\infty}^t R_{a_j}(t-\tau)\langle s(\tau) \rangle d\tau,
\label{eq:Ra-at}
\end{equation}
with $\langle \cdot\rangle$ and $\langle \cdot\rangle_u$ denoting noise average with and without an external time-varying signal, respectively. Without loss of generality, we set the stationary activity $\langle a_j\rangle_{u}$ to zero. 
{\color{black} The ``activity response function'' $R_a$ is a property of the intracellular molecular network,  which can be computed for specific models~\cite{Wang2016entropy,wang2018inferring} or measured directly in single-cell experiments~\cite{martin2001comparison,Shimizu2010EcoliResponse,sgro2015intracellular,kamino2017fold}.  
In general,  $R_a$ may depend on the ambient signal level $s$ of the cell. 

The shared signal $s$ offers a means to synchronise the activities of cells. We derive here a matching condition
for $s$ and the $a$'s to enter a positive signal relay. Expressing Eq.~(\ref{eq:Ra-at}) in Fourier form, we have 
$\langle\tilde{a}_j(\omega)\rangle=\tilde{R}_{a_j}(\omega)\langle \tilde{s}(\omega)\rangle$.   
For weak disturbances, the restoring force in Eq.~(\ref{eq:general-adaptation-4})
can be approximated by a linear one, i.e., $F( s)\simeq-Ks$. 
Consequently, $\langle \tilde{s}(\omega)\rangle=\sum_{j=1}^N\alpha_1 \tilde{R}_s(\omega)\langle\tilde{a}_j(\omega)\rangle$, 
where 
\begin{equation}
\tilde{R}_s=\frac{1}{K-i\gamma\omega}
\label{eq:Rs}
\end{equation}
is the ``signal response function'', with $i$ the imaginary unit.  
For identical cells, these equations yield an oscillatory solution
$\tilde{a}(\omega_o)\neq 0$ provided $N\alpha_1\tilde{R}_a(\omega_o)\tilde{R}_s(\omega_o)=1$.   
To gain more insight, we express the two response functions in their amplitudes and phase shifts, i.e. 
$\tilde{R}_a\equiv |\tilde{R}_a|\exp(-i\phi_a)$ and $\tilde{R}_s\equiv |\tilde{R}_s|\exp(-i\phi_s)$.  
Then, the cell density $N=N_o$ and the selected frequency $\omega_o$ at the onset of collective oscillations are determined by,
\begin{subequations}\label{eq:self-consistency}
\begin{eqnarray}
\label{eq:self-consistency-phase}
\phi_a(\omega_o)&=&-\phi_s(\omega_o),\\
\label{eq:self-consistency-amplitude}
|\tilde{R}_a(\omega_o)\tilde{R}_s(\omega_o)|&=&(\alpha_1N_o)^{-1}.
\end{eqnarray}
\end{subequations}
These are essentially conditions of linear instability for the quiescent state expressed in terms of the single-cell and signal response
functions, and constitute our first main result. For inhomogeneous cell populations, one simply replaces $R_a$
by its population average $\bar{R}_a\equiv N^{-1}\sum_{j=1}^N R_{a_j}$.

Under the very general assumption of additive signal release from individual cells as
expressed by Eq.~(\ref{eq:general-adaptation-4}),  we now have a mathematical prediction for the onset density 
$N_o$ and oscillation frequency $\omega_o$.
Let $\alpha_2\sim |\tilde{R}_a|$ be the sensitivity of the cell activity against $s$. We introduce a
``signal relay efficiency'' $\bar{N}\equiv  N\alpha_1\alpha_2$, which also sets the coupling strength of cellular activities
through the signal.   Oscillations start at the critical coupling strength $\bar{N}_o=N_o\alpha_1\alpha_2$.  Eq.~(\ref{eq:self-consistency-amplitude}) simply states that, at the selected frequency $\omega_o$, signal amplification through the 
collective action of $N_o$ cells compensates signal loss from dissipative forces acting on $s$, 
e.g., friction for a mechanical signal or degradation/dilution for a chemical signal.  
The frequency $\omega_o$ is chosen such that  phase shifts incurred in the forward and reverse medium-cell transmissions
match each other [Eq.~(\ref{eq:self-consistency-phase})].  
\\ 

\noindent
\textbf{Cell-to-signal energy flow.} 
Auto-induced collective oscillations must be driven by intracellular active processes.  
These active components of the system give a nonequilibrium character to the activity response~\cite{lan2012energy,cao2015oscillation,Sartori2013Kinetic,shouwen2015adaptation}
and furthermore enable energy flow from the cell to the signal upon periodic stimulation, an interesting physical phenomenon 
left unnoticed so far. }

To set the stage, we turn to basic considerations of non-equilibrium thermodynamics~\cite{sekimoto2010stochastic,seifert2012stochastic}. 
The shared signal $s$ as illustrated in Fig.~\ref{fig:scenarios} typically follows a dissipative dynamics such as 
Eq.~(\ref{eq:general-adaptation-4}). When the medium is close to thermal equilibrium, the Fluctuation-Dissipation Theorem (FDT) 
relates the imaginary component $\tilde{R}_s''$ of the signal response $\tilde{R}_s$ to its spontaneous fluctuation $\tilde{C}_s$ 
induced by thermal noise~\cite{kubo1966fluctuation,Wang2016entropy,wang2018inferring}: 
$2T\tilde{R}_s''(\omega)=\omega \tilde{C}_s(\omega)$, 
where $\tilde{C}_s(\omega)=\langle |\tilde{s}(\omega)|^2\rangle_{u}$ is the spectral amplitude of the signal, and $T$ is the temperature.  
This relation demands $\tilde{R}_s''(\omega)$ to be positive at all frequencies. Hence, the dissipative nature of the physical environment translates into a phase delay, i.e., $\phi_s\equiv -\arg (\tilde{R}_s)\in  (-\pi,0)$.  
{\color{black}  Under the over-damped signal dynamics [Eq.~(\ref{eq:general-adaptation-4})],  Eq.~(\ref{eq:Rs}) gives
 \begin{equation}
 \phi_s(\omega)\equiv -\arg\Big(\tilde{R}_s(\omega)\Big)=-\tan^{-1}(\omega \tau_s)\in \bigl(-{\pi\over 2},0\bigr),
 \label{eq:phi_s}
 \end{equation}
where $\tau_s=\gamma/K$ is the signal relaxation time.
(The situation $-\pi<\phi_s(\omega)<-\pi/2$ occurs at high frequencies when the dynamics of $s$ is underdamped.)
On the other hand, a leading phase as required by Eq.~(\ref{eq:self-consistency-phase}) for the intracellular signal relay, 
violates the FDT. In the present case, active cells play the role of the out-of-equilibrium partner. 
We have calculated the work done by one of the cells on the signal when the latter oscillates at a frequency 
$\omega$ (Supplementary \rm{II}). 
The output power $\dot{W}\equiv \langle \dot{s}\cdot \alpha_1 a\rangle$, i.e., the averaged value of
the product between signal velocity  ($\dot{s}$) and force from an individual cell ($\alpha_1a$), is given by 
\begin{eqnarray}
\dot{W}&\simeq& -\alpha_1\omega \tilde{R}_a''(\omega)\langle|\tilde s(\omega)|^2\rangle\nonumber\\
&=&\alpha_1\omega |\tilde{R}_a(\omega)|\sin\phi_a(\omega)\langle|\tilde s(\omega)|^2\rangle.
\label{eq:energy-release}
\end{eqnarray}
The energy flux is positive, i.e., flowing from the cell to the signal, when $a$ has a phase lead over $s$,
re-affirming Eq.~(\ref{eq:self-consistency-phase}) as a necessary condition on thermodynamic grounds.    
Stimulated energy release from an active cell to the signal as expressed by
Eq.~(\ref{eq:energy-release}) constitutes our second main result in this paper. 

Eq.~(\ref{eq:energy-release}) can also be used to calculate the energy flux for an arbitrary signal time series $s(t)$, provided
the linear response formula Eq.~(\ref{eq:Ra-at}) applies. In particular, thermal fluctuations of $s$ in
the quiescent state may activate a net cell-to-signal energy flow. 
The total power is obtained by integrating contributions from all frequencies. }
Previous experiments from Hudspeth lab yielded a phase-leading response of hair bundles to mechanical stimulation at
low frequencies~\cite{martin2001comparison}. The same group also showed
that energy can be extracted from the hair bundle via a slowly oscillating stimulus~\cite{Martin1999hairResponse}.  \\

 \noindent
\textbf{Chemical oscillations.} 
The criteria given by Eq.~(\ref{eq:self-consistency}) apply equally to chemical oscillations illustrated in Fig.~\ref{fig:scenarios}{\it B}.  
In contrast to the mechanical system, Eq.~(\ref{eq:general-adaptation-4}) at $\gamma = 1$ becomes a rate equation for
the extracellular concentration $s$ of the signalling molecules.
The term $F(s)$ (negative) gives the degradation or dilution rate of $s$ in the medium, while individual cells secrete the molecules
at a rate proportional to their activity $a$. 
As the signalling molecules are constantly produced and degraded, chemical equilibrium is often violated even in the steady state.
Nevertheless, $F(s)$ usually plays the role of a stabilising force so that the signal response function
$\tilde{R}_s(\omega)$ has the same phase-lag behaviour as the mechanical case.
Release of the molecules by the communicating cells must be phase-leading so as to drive oscillatory signalling.  
\\

\noindent
{\color{black}
\textbf{Adaptive cells show phase-leading response.} 
Apart from the aforementioned hair bundles, phase-leading response to a low frequency signal has also been reported in the activity 
of \emph{E. coli} chemoreceptors~\cite{Shimizu2010EcoliResponse} and in the osmo-response in yeast~\cite{mettetal2008frequency}.  Interestingly, all three of these cases are examples of adaptive sensory systems whose response to a step signal at $t=0$ is shown in
Fig.~\ref{fig:adaptation_illustration}{\it A}. The small activity shift $\epsilon$ at long times is known as the ``adaptation error''.
Fig.~\ref{fig:adaptation_illustration}{\it B} shows the response of the same system under a sinusoidal signal.
The low frequency response exhibits a phase lead while the high frequency one has a phase lag.
Below, we show that the sign switch in the phase shift of an adaptive variable is an inevitable consequence of causality.

\begin{figure}
\centering
\includegraphics[width=8.5cm]{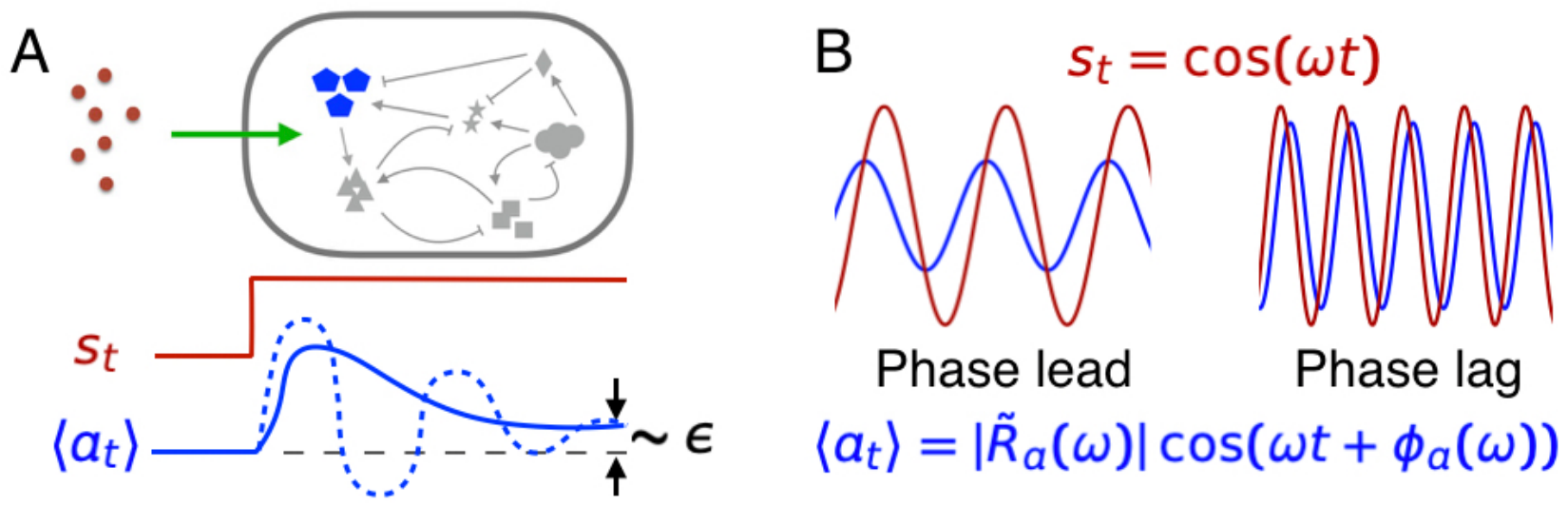}
\caption{Dynamical response of an intracellular adaptive variable $a$.
({\it A})  Response to a stepwise signal:  after a transient response, $a$ returns to its 
pre-stimulus state (within a small error $\epsilon$). 
In the simplest case, the transient response is controlled by the activity shift timescale $\tau_a$ and the circuit feedback 
timescale $\tau_y$. Solid and dashed lines correspond to over-damped ($\tau_a\ll\tau_y$) and 
under-damped ($\tau_a\gg\tau_y$) situations, respectively.   
({\it B})  Response to a sinusoidal signal at low (left) and high (right) frequencies.  
The phase shift $\phi_a$ switches sign. }
\label{fig:adaptation_illustration}
\end{figure}

From the causality condition $R_a(t<0)=0$, the real ($\tilde{R}_a'$) and imaginary ($\tilde{R}_a''$) part 
of the response function in frequency space satisfy the Kramers-Kr\"onig relation~\cite{Sethna2006entropy}: 
\begin{equation}
\tilde{R}_a'(\omega)=\frac{2}{\pi} \int_0^\infty \tilde{R}_a''(\omega_1)\frac{\omega_1}{\omega_1^2-\omega^2}d\omega_1.
\label{eq:kramers}
\end{equation}
For a step signal of unit strength, Eq.~(\ref{eq:Ra-at}) yields
\begin{equation}\label{eq:adaptation-error}
\epsilon=\langle a(\infty)\rangle-\langle a\rangle_u= \int_0^\infty R_a(\tau) d\tau =\lim_{\omega\rightarrow 0}\tilde{R}'_a(\omega).
\end{equation} 
Comparing Eqs.~(\ref{eq:kramers}) and (\ref{eq:adaptation-error}) in the limit $\omega\rightarrow 0$ and assuming $\epsilon$
to be sufficiently small, we see that $\tilde{R}_a''(\omega)$ inside the integral must change sign.
In other words, both phase-leading ($\tilde{R}_a''<0$) and lagged  ($\tilde{R}_a''>0$) behaviour are present across the frequency domain. {\color{black} This is our third result. }

Adaptation plays a key role in biochemical networks~\cite{goentoro2009evidence,cohen2009dynamics}, 
and especially in sensory systems~\cite{alon1999robustness,reisert2001response,hohmann2002osmotic,hazelbauer2008bacterial,menini1999calcium,nakatani1991light,mettetal2008frequency,hudspeth2014integrating}. 
Connection between adaptation and collective oscillations has been
implicated in previous  works~\cite{kamino2017fold, matsuoka1985sustained,Mello2004-tn}. 
With the mathematical results presented above, the logical link between adaptation and phase-leading response, and onto collective oscillations through signal relay, is firmly established.  Below we illustrate details of this process in
three adaptive systems of increasing complexity. Implications of our model study to experimental work are given in the
Discussion section.\\

\noindent

\textbf{A weakly nonlinear model with adaptation.} 
 We consider first a noisy two-component circuit which is a variant of the model 
for sensory adaptation in {\it E. coli}~\cite{lan2012energy} (Fig.~\ref{fig:weakly-adaptive-model}{\it A}, see also Methods). 
The receptor response function in the quiescent state is given by
\begin{equation}
\tilde{R}_a(\omega)= \alpha_2 \Bigl[1+{\epsilon\over \epsilon^2+(\tau_y\omega)^2}+i\tau_a \omega^\ast 
{(\omega^\ast /\omega)-(\omega/\omega^\ast )\over 1+\bigl(\epsilon/(\tau_y\omega)\bigr)^2}\Bigr]^{-1},
\label{eq:linear-R}
\end{equation}
where 
\begin{equation}
\omega^\ast=(\tau_a\tau_y)^{-1/2}(1-\epsilon^2\tau_a/\tau_y)^{1/2}.
\label{eq:omega_max}
\end{equation}
Here $\tau_a$ and $\tau_y$ are the timescales for the activity ($a$) and negative feedback ($y$) dynamics, respectively.
In Figs.~\ref{fig:weakly-adaptive-model}{\it B,C}, we show the phase shift $\phi_a(\omega)$ and
the real and imaginary part of $\tilde{R}_a(\omega)$ against the frequency $\omega$, plotted on semi-log scale.
As predicted, $\phi_a(\omega)$ undergoes a sign change at $\omega^\ast $. Correspondingly, the imaginary component of the 
response $\tilde{R}_a''$ becomes negative in the phase-leading regime, violating the FDT. 
The peak of $|\tilde{R}_a(\omega)|$ is located close to $\omega^\ast$, with a relative width 
$\Delta\omega/\omega^\ast\simeq Q^{-1}$ where $Q=\tau_a\omega^\ast\simeq (\tau_a/\tau_y)^{1/2}$.

Allowing the chemoreceptor activity $a$ to affect the signal as in Eq.~(\ref{eq:general-adaptation-4})
with $F(s)=-Ks$, we observe an oscillatory phase upon increase in cell density in numerical simulations (Fig.~\ref{fig:weakly-adaptive-model}{\it D}).
Fig.~\ref{fig:weakly-adaptive-model}{\it E} shows the oscillation amplitude (upper panel) and frequency (lower panel) against
the coupling strength $\bar{N}$ around the onset of oscillations.
The threshold coupling strength $\bar{N}_o=N_o\alpha_1\alpha_2$ 
and the onset frequency $\omega_o$ both agree well with the values
predicted by Eq.~(\ref{eq:self-consistency}) (see arrows in Fig.~\ref{fig:weakly-adaptive-model}{\it E}).
The transition is well described by a supercritical Hopf bifurcation.
At finite oscillation amplitudes, there is a downward shift of the oscillation frequency which can be 
quantitatively calculated in the present case by introducing a renormalised response function $\tilde{R}_a^+(\omega)$ whose
phase is shown in Fig.~\ref{fig:weakly-adaptive-model}{\it F} (see Supplementary \rm{III}).   
The oscillation frequency is determined by the crossing of the two curves $\phi_s(\omega)$ and $\phi_a^+(\omega)$,
with the formal independent of the oscillation amplitude $A$.
As the oscillation amplitude grows further, higher order harmonics generated by the nonlinear term become more prominent. 
The coupled system eventually exits from the limit cycle through an
infinite-period bifurcation and arrives at a new quiescent state.  
The upper bifurcation point $\bar{N}_b$ is inversely proportional to the adaptation error $\epsilon$ (see Supplementary Fig.~S1).

\begin{figure}
\centering
\includegraphics[width=8.5cm]{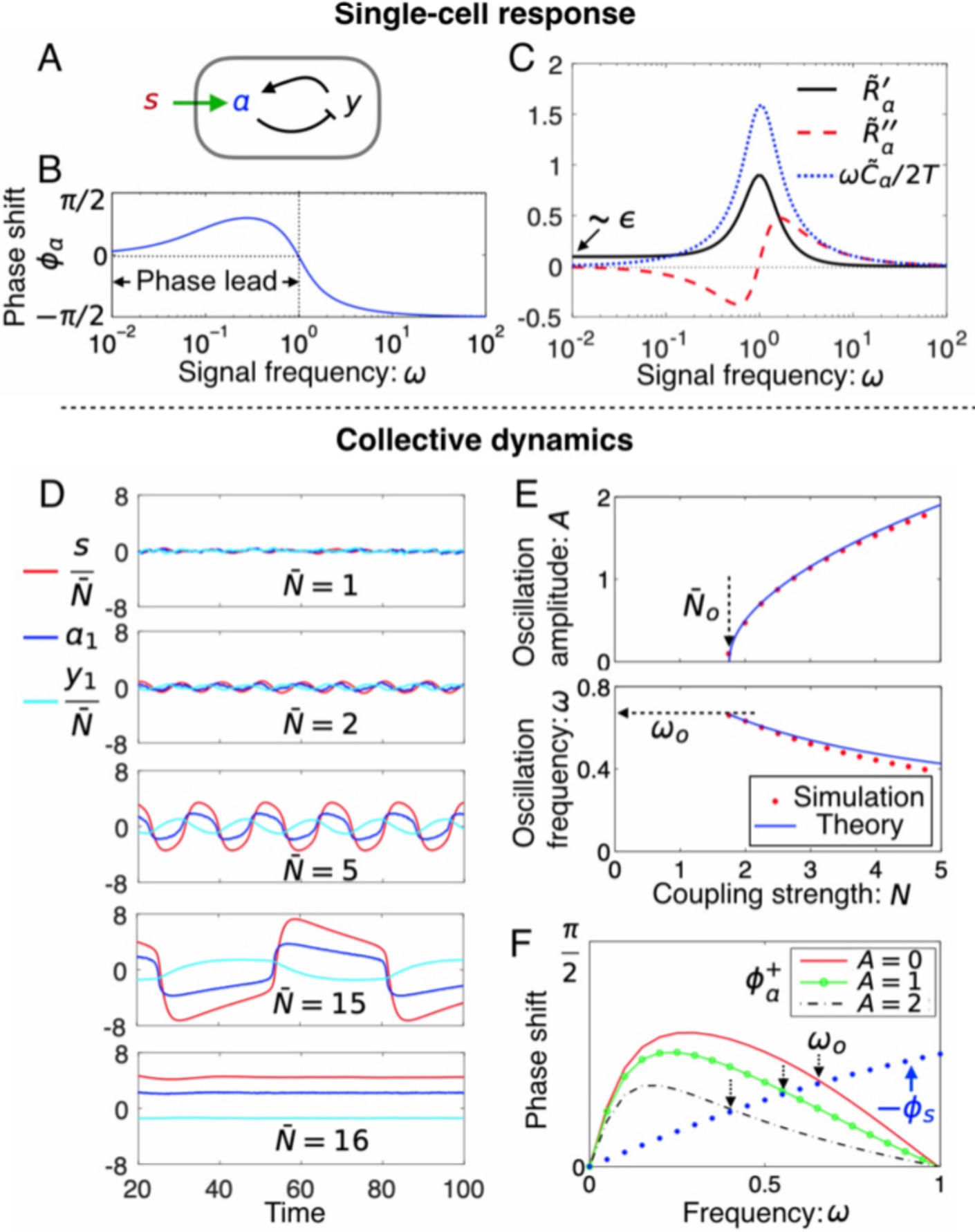}
\caption{A weakly nonlinear model with adaptation.  ({\it A-C}) Single cell response.  
({\it A})  A noisy two-component model with negative feedback. 
({\it B})  Frequency-resolved phase shift $\phi_a=-\arg(\tilde{R}_a)$. A sign change takes
place at $\omega=\omega^\ast $, with $a$ leading $s$ on the low frequency side. 
({\it C}) Real ($\tilde{R}_a'$) and imaginary ($\tilde{R}_a''$) components of the response spectrum.  
$\tilde{R}_a'$ is of order $\epsilon$ in the zero frequency limit, while $\tilde{R}_a''$ changes sign at $\omega=\omega^\ast $.   
Also shown is the correlation spectrum $\tilde{C}_a(\omega)$ multiplied by $\omega/(2T)$, where $T$ is the noise strength.
The fluctuation-dissipation theorem $\tilde{R}_a''=\omega\tilde{C}_a(\omega)/(2T)$ for thermal equilibrium
systems is satisfied on the high frequency side, but violated at low frequencies. 
({\it D-F}) Simulations of coupled adaptive circuits.
({\it D}) Time traces of the signal (red) and of the activity (blue) and memory (cyan) from one of 
the participating cells at various values of the coupling strength $\bar{N}=\alpha_1\alpha_2N$.     
({\it E})  The oscillation amplitude $A$ (of activity $a$) and frequency $\omega$ against $\bar{N}$.  The amplitude $A$ grows as $(\bar{N}-\bar{N}_o)^{1/2}$ here, a signature of Hopf bifurcation.  ({\it F}) Determination of oscillation frequency from the renormalised phase matching condition 
at finite oscillation amplitudes:  $\phi_a^+(\omega,A)=-\phi_s^+(\omega,A)$.  The linear model for $s$  
yields $\phi_s^+(\omega,A)=-\phi_s(\omega)$. 
Parameters: $\tau_a=\tau_y=\gamma=K=c_3=1$, $\alpha_1=\alpha_2=0.5$, and $\epsilon=0.1$. 
The strength of noise terms is set at $T=0.01$.
}
\label{fig:weakly-adaptive-model}
\end{figure}

The signal phase shift $\phi_s(\omega)$ is given by Eq.~(\ref{eq:phi_s}).
When the signal relaxation time $\tau_s$ is much shorter than the cell adaptation time
$\tau^*\equiv 2\pi/\omega^\ast $, $\phi_s(\omega)$ stays close to zero so that the selected 
period is essentially given by $\tau^*$. In this case, $|\tilde R_a(\omega)|$ is near its peak and hence the cell density
required by Eq.~(\ref{eq:self-consistency-amplitude}) is the lowest.
As signal clearance slows down, the crossing point shifts to lower frequencies. Given a finite adaptation
error $\epsilon>0$, there is a  generic maximum signal relaxation time $\tau_s^*\sim\epsilon^{-1}$
beyond which the phase matching cannot be achieved (see Supplementary {\rm III. C}).\\

\noindent
\textbf{Excitable dynamics.} 
 DQS in {\it Dictyostelium} and other eukaryotic cells takes the form of pulsed release of signalling molecules~\cite{gregor2010onset,hubaud2017excitable,tang2014evolutionarily}. 
The highly nonlinear two-component FitzHugh-Nagumo (FHN) model is often employed for such excitable phenomena~\cite{lindner2004effects,izhikevich2007dynamical,sgro2015intracellular}. 
Similar to the sensory adaptation model discussed above, each FHN circuit has a 
memory node $y$ that keeps its activity $a$ low (the ``resting state'') under a slow-varying 
signal $s(t)$ (Fig.~\ref{fig:FHN_model}{\it A}, see also Methods).
On the other hand, a sufficiently strong noise fluctuation or a sudden shift of $s$ sends the circuit through a large excursion 
in phase space (known as a ``firing event'') when $y$ is slow (i.e., $\tau_y\gg\tau_a$). 
Our numerical investigations show that firing does not disrupt the adaptive nature of the circuit under the negative feedback from $y$.
The noise-averaged response of a single FHN circuit exhibits the same characteristics as the sensory adaptation model,
including adaptation to a stepwise stimulus after a transient response (Fig.~\ref{fig:FHN_model}{\it B}, upper panel), 
and the phase-leading behaviour and diminishing response amplitude on the low frequency side (Fig.~\ref{fig:FHN_model}{\it B}, lower panel).

\begin{figure}[t]
\centering
\includegraphics[width=8.5cm]{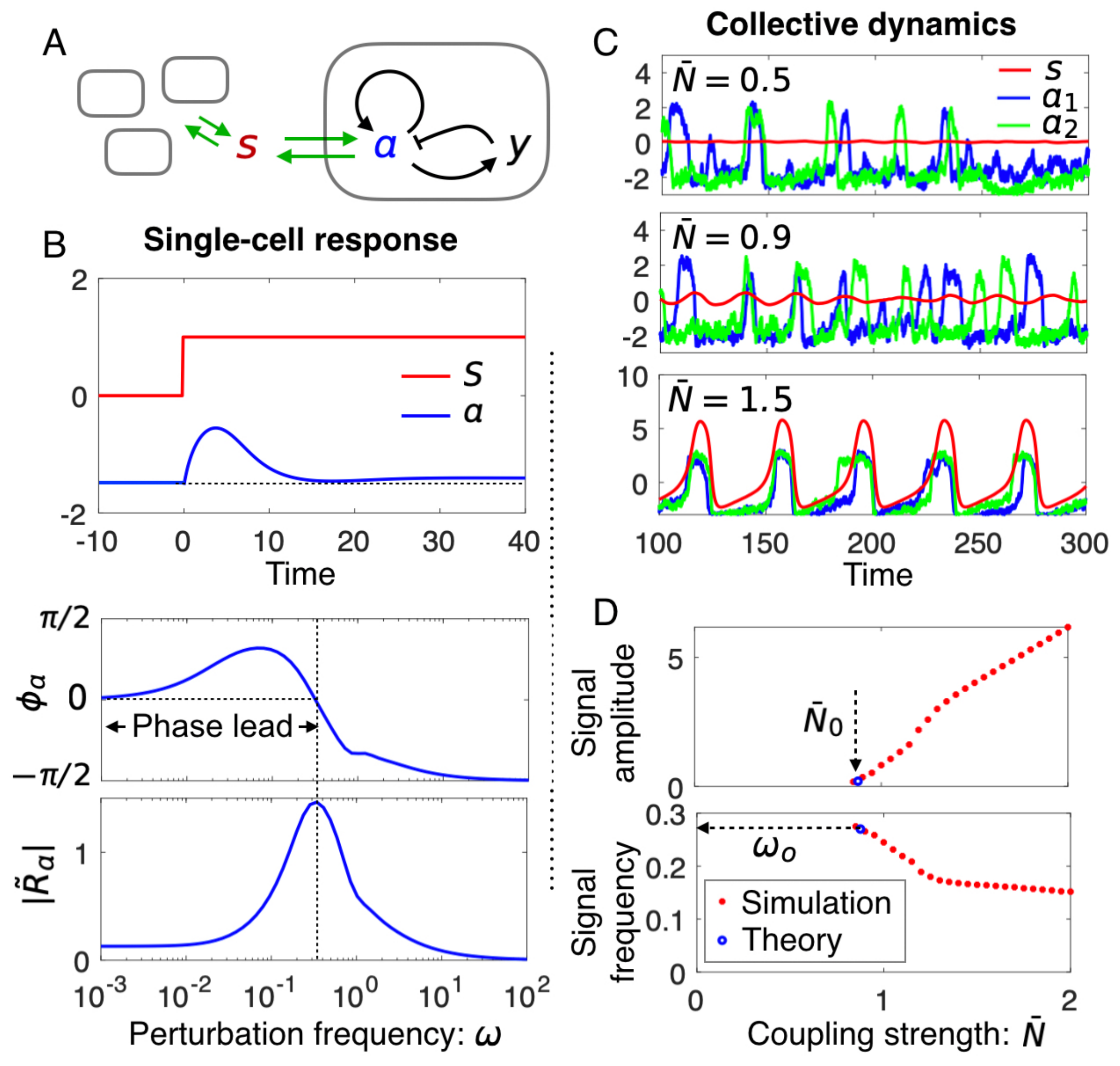}
\caption{Simulations of the coupled excitable FitzHugn-Nagumo (FNH) model with noise. ({\it A}) Model illustration. 
Note the self-activation of $a$ that gives rise to excitability (see Methods for details).  ({\it B}) Noise-averaged response of $a$ in the resting state.  Upper panel:  the average response to a step signal.  Lower panel:  the response amplitude and phase shift at various signal frequencies.    ({\it C}) Trajectories of the coupled FHN model at various values of the effective coupling strength $\bar{N}$.   
In addition to the signal $s$, activities of two out of a total of 1000 cells are plotted.    ({\it D})  Signal oscillation amplitude
and frequency against effective cell density. Red stars: simulation data; Blue circles: predictions 
of Eq.~(\ref{eq:self-consistency}) using numerically computed response spectra. }
\label{fig:FHN_model}
\end{figure}

Fig.~\ref{fig:FHN_model}{\it C} shows time traces of individual cell activities (blue and green curves) as well as that of the signal 
$s$ (red curve) from simulations of weakly coupled FHN circuits at three different values of the coupling strength
$\bar{N}$ (see Methods). At $\bar{N}=0.5$, the two selected cells fire asynchronously while $s$ remains constant.
At $\bar{N}=0.9$, collective behaviour as seen in the oscillation of $s$ starts to emerge,
although individual circuits continue to fire sporadically. Upon further increase of $\bar{N}$, synchronised firing is seen. 
Despite the highly nonlinear nature of the FHN model, both the onset coupling strength $\bar{N}_o$ and the frequency $\omega_o$
are well predicted by Eq.~(\ref{eq:self-consistency}) using the respective response functions in the resting state
(Fig.~\ref{fig:FHN_model}{\it D}). \\

\noindent
\textbf{Yeast glycolytic oscillations: an oscillator imbedded in a complex biochemical network.}
We take the adapt-to-oscillate scenario one step further to examine the dynamics of ATP autocatalysis in yeast. 
Concentration oscillations of NADH and glycolytic intermediates have been observed in yeast cell extracts as well as 
in starved yeast cell suspensions upon shutting down the respiratory pathway (see Ref. \cite{richard2003rhythm} for a review). 
The phosphofructokinase (PFK), an enzyme in the upper part of the glycolytic pathway, is tightly regulated by ATP,  
which is also a key product of glycolysis. 
This robust negative feedback is commonly regarded as the driver of glycolytic oscillations, 
with a typical period of 30-40 seconds in intact cells but 2 minutes or longer in extracts. 
Cells at high density show collective oscillations due to cell-to-cell communication via the freely diffusing molecule 
acetaldehyde (ACE)~\cite{richard2003rhythm,amemiya2015collective}. 
As the cell density decreases, the synchronised behaviour breaks down. While many studies found continued oscillation
of individual cells at their own frequencies~\cite{gustavsson2012sustained,weber2012desynchronisation}, simultaneous disappearance of individual and collective oscillations as in other DQS systems
has also been reported\cite{de2007dynamical}.
In the following, we present results of a detailed model study that yield quantitative insights on the underlying
intracellular biochemical network and how it interfaces with the diffusing chemical ACE.

Our starting point is the du Preez {\it et al.} model of yeast glycolysis and fermentation for a single cell~\cite{du2012steady}. 
It includes around 20 metabolic reactions (Fig.~\ref{fig:glycolysis}{\it A}).
The extracellular environment is set by glucose and ACE concentrations.
Simulations of the model under steady environmental conditions yields a phase diagram shown in Fig.~\ref{fig:glycolysis}{\it B}, 
with coloured regions showing steady metabolic flow and the white regime spontaneous oscillations~\cite{gustavsson2015entrainment}. 
The glucose concentration, which controls glycolytic flux, needs to be sufficiently high for oscillations to take place.
ACE also has a role in the dynamics: either very low or very high concentrations arrest the oscillations.

We now examine the response of intracellular metabolites to a sudden shift in the extracellular ACE concentration.
Fig.~\ref{fig:glycolysis}{\it C} gives the concentration variation of four metabolites at three selected points on the left side of
the phase diagram. In all cases, NAD follows closely ACE concentration change and hence acts as an instantaneous transducer 
of the signal. ATP adapts best while PYR, the substrate to produce ACE, adapts less accurately.
TRIO, the metabolite immediately upstream of the enzyme GAPDH that uses NAD as cofactor, does not adapt.
Interestingly, the response of ATP and of TRIO switches sign at ACE$_0\simeq 0.2$ mM, where the tip of the white regime is located.
Overall, the adaptation error increases progressively as one moves away from the oscillatory regime.
The graded adaptation error is illustrated in Fig.~\ref{fig:glycolysis}{\it B}, where PYR adaptation is limited to the blue regime
while ATP adaptation covers both blue and orange regions.

Fig.~\ref{fig:glycolysis}{\it D} shows phase shifts of ATP, NAD and five other metabolites along the glycolytic pathway
to a periodic ACE signal at various frequencies. In the upper panel, which corresponds to the point marked by star on Fig.~\ref{fig:glycolysis}{\it B}, ATP, BPG and PEP have a leading phase (after a $\pi$ shift) below the frequency $20$ min$^{-1}$.  
The list is expanded to all six metabolites in the lower panel (diamond in the blue regime), including PYR
which is directly upstream of the signalling molecule ACE. Cells in the blue part of the phase diagram are
susceptible to adaptation-driven oscillations according to our theory. 

Our simulation studies of the model indicate that the glyoxylate shunt (GLYO), which consumes ATP, plays an important role in
modulating the effect of ACE on the glycolytic flux. Flux through GLYO is low at low ACE concentrations. 
In this case, an increase in ACE concentration elevates the NAD/NADH ratio which in turn pushes up the flux through GAPDH. 
This is evidenced in the rise of ATP and PYR concentrations immediately after an upshift of ACE  
(Fig.~\ref{fig:glycolysis}{\it C}, bottom plot).
As ACE concentration increases, GLYO flux increases as well, leading to a drop in ATP concentration.
The opposing effects of GAPDH and GLYO reactions are behind the sign reversal in the response of ATP against ACE upshift.
On the dashed line in Fig.~\ref{fig:glycolysis}{\it B} at ACE$_0\simeq 0.2$ mM, the two effects cancel each other, rendering ATP irresponsive to the signal. 
Artificially turning off the GLYO reaction, we extend the positive transient response of ATP from
below the dashed line to the whole phase diagram (Supplementary S10-S11). In this case, the adaptive regimes of ATP and PYR 
become identical to each other, so that the blue regime on the phase diagram is greatly expanded (Supplementary Fig.~S10).  

\begin{figure}[t]
\centering
\includegraphics[width=8.5cm]{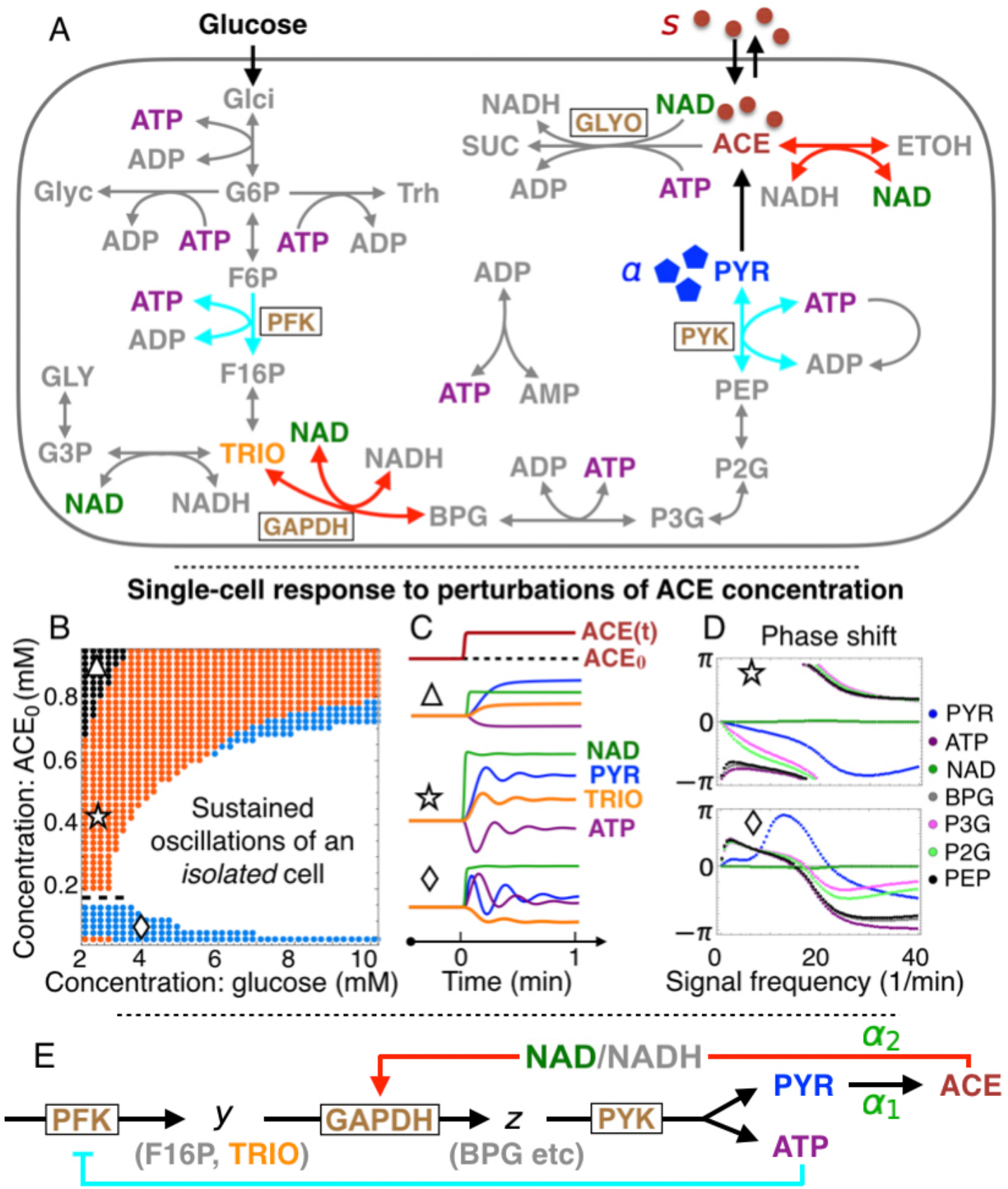}
\caption{Yeast glycolytic oscillations. ({\it A}) The reaction network of glycolysis in a yeast cell 
(see Supplementary Fig.~S3 for full names of the abbreviations).   
({\it B}) Single-cell phase diagram spanned by the extracellular glucose and acetaldehyde (ACE) concentrations.
Coloured regions illustrate graded adaptation of metabolites to an upshift in ACE, with representative time traces given
in ({\it C}). The response of ATP changes sign around ACE$_0\simeq0.2$ as indicated by the dashed line. 
({\it D}) Frequency-resolved phase shifts of selected metabolites to weak sinusoidal perturbations. 
ATP, BPG and PEP are phase-leading in both blue and orange regions of the phase diagram, while PYR does so only in the
blue regime. ({\it E})  A reduced model for glycolytic oscillations where the intracellular NAD/NADH ratio and 
pyruvate (PYR) act as the receiver and sender of the signal (ACE), respectively. 
Adaptive response of PYR to ACE is coupled to the homeostasis of ATP through the reaction PYK.  } 
\label{fig:glycolysis}
\end{figure}

To further understand these dynamical properties, we constructed a reduced model in Fig.~\ref{fig:glycolysis}{\it E} 
by taking into account stoichiometry and known regulatory interactions along the glycolytic pathway~\cite{chandra2011glycolytic}, 
and by making use of the timescale separation in the turnover of metabolites as suggested by their response spectra (Fig.~\ref{fig:glycolysis}{\it D} and Supplementary Fig.~S8).  
Since ATP and PYR now appear as co-products of the condensed reaction PYK in the reduced model, the latter can be viewed
as a ``reporter'' of ATP homeostasis implemented by the negative feedback loop (cyan in Fig.~\ref{fig:glycolysis}{\it E}). 
We have verified that the reduced model exhibits a very similar response to ACE signal as the full model, 
demonstrating insensitivity of ATP homeostasis-mediated adaptation to model details
(see Supplementary Fig.~S9 and also Supplementary Fig.~S11 when the GLYO reaction is turned off).

We have also investigated collective behaviour in a coupled system where individual cells metabolise according to 
the reduced model and communicate their internal NAD/NADH ratio through the shared signal ACE.
For identical cells, the system enters the collectively oscillating state on the low density side when the ACE level,
produced by cells themselves, crosses into the oscillatory regime on the phase diagram (Supplementary Fig. S9{\it D}).
At the normalised cell density $\phi_c=0.34$, the ACE level reaches $s_c=0.72$ (arbitrary unit), where an isolated cell
exits from the oscillatory regime into the adaptive regime. The coupled cell population, on the other hand, continues
to oscillate through DQS (Supplementary Figs.~S12-S14).   
Although DQS represents a conceptually different scenario from the Kuramoto model of oscillator synchronisation,
the two merge seamlessly in the present case. The expanded range of external conditions (i.e., blue and white regions of the phase diagram) contributes to the robustness of collective oscillations over a broader range of cell densities, 
particularly when the glyoxylate shunt is turned off. \\

\noindent
{\large \textbf{Discussion}}\\
In this work, we investigated a general scenario for emerging oscillations in a group of cells that communicate via a shared signal. 
It covers a broad class of pulsation behaviour in cell populations, collectively known as dynamical quorum sensing. 
Using the single-cell response to external stimulation, we formulated a quantitative requirement for the onset of collective oscillations
that must be satisfied by active cells as well as models of them. 
A proof is presented to link this requirement to the adaptive release of signalling molecules by individual cells. 
Our work thus consolidates observations made in the literature and formalises adaptation as a unifying theme behind DQS. 

The above mathematical results connect well to the recent surge of interest in active systems, where collective phenomena 
emerge due to energy-driven processes on the microscopic scale~\cite{chen2017weak,kawaguchi2017topological}. 
The study of such non-equilibrium processes opens a new avenue to explore 
mechanisms of spontaneous motion on large scales.
We presented a general formula for the energy outflow of a living cell through a designated mechanical or chemical channel
under periodic stimulation. This energy flux is positive over a range of frequencies when the cell responds to the stimulus adaptively. 
Since adaptation is a measurable property of a cell, the thermodynamic relation is applicable without making specific 
assumptions about intracellular biochemical and regulatory processes, while most models do.
When cells are placed together in a fixed volume, a quorum is required to activate the energy flow via self and mutual stimulation.

We reported three case studies to illustrate how these general yet quantitative relations could be applied to 
analyse the onset of collective oscillations in specific cellular systems.
Our first example is a coarse-grained model where signal reception and release are integrated 
into the same activity node (e.g., a membrane protein or a molecular motor). Due to the weak nonlinearity of the intracellular circuit, 
many analytical results were obtained. The intracellular adaptive circuit has two timescales:
the activity relaxation time $\tau_a$ and the negative feedback time $\tau_y$. Their ratio $Q^2=\tau_a/\tau_y$,
similar to the quality factor in resonators, determines the shape of the adaptive response (Fig. 2). 
At small adaptation error $\epsilon\ll 1$, the imaginary part of the response function $\tilde{R}_a(\omega)$ changes sign at the characteristic frequency $\omega^\ast\simeq (\tau_a\tau_y)^{-1/2}$.
This is also approximately the frequency where $|\tilde{R}_a(\omega)|$ reaches its maximum.
When cells are coupled through the signal with a relaxation time $\tau_s$, the onset oscillation frequency $\omega_o$ increases
with decreasing $\tau_s$, reaching its maximal value $\omega^\ast$ when $\tau_s\ll 1/\omega^\ast$.

Much of these results carry over to our second example, a population of coupled excitable circuits described by the 
FitzHugh-Nagumo model.  Despite its highly nonlinear nature, the FHN model in the resting state shows adaptive 
response under weak stimulation. Our numerical simulations of the coupled system at weak noise confirm the 
onset oscillation frequency and the critical cell density predicted by Eq.~(\ref{eq:self-consistency}). 

The above quantitative predictions compare favourably with available experimental data.  
The first is mechanical stimulation of hair cells carried out by Martin {\it et al.}~\cite{martin2001comparison}, 
where the cellular response was extracted using a flexible glass fibre.  
Deformation of the glass fibre, which is the signal here, has a relaxation timescale ($\sim 0.5$ ms) much shorter 
than the adaptation time of the hair bundle ($\sim 0.1$ s).  Spontaneous oscillations of the combined system were
observed at 8 Hz, the predicted frequency where the imaginary part of the hair bundle response function $\tilde{R}_a''(\omega)$
undergoes the expected sign change.  The second is a recent microfluidic single-cell measurement of  
{\it Dictyostelium} reported by Sgro {\it et al.}~\cite{sgro2015intracellular}, where the change of cytosolic cAMP level (activity $a$) in response to extracellular cAMP variation (signal $s$) was presented.  From the measured response $a(t)$ to a step increase 
of the signal in their work (reproduced in Fig.~\ref{fig:Dicty-phase}{\it A, Upper panel}),  we computationally deduced the 
response function $R_a(t)=da/dt$ in the time domain (Fig.~\ref{fig:Dicty-phase}{\it A, Lower panel}) and then the response spectrum 
$\tilde{R}_a$ via Fourier transform.  The resulting phase shift $\phi_a$ changes sign around 
$\omega^\ast=1$ min$^{-1}$ (Fig.~\ref{fig:Dicty-phase}{\it B}).
 According to our theory, the onset oscillation period at high flow rates should be around 6.28 min, which is indeed what was
observed in experiments~\cite{gregor2010onset,sgro2015intracellular,kamino2017fold}.

\begin{figure}[t]
\centering
\includegraphics[width=8.5cm]{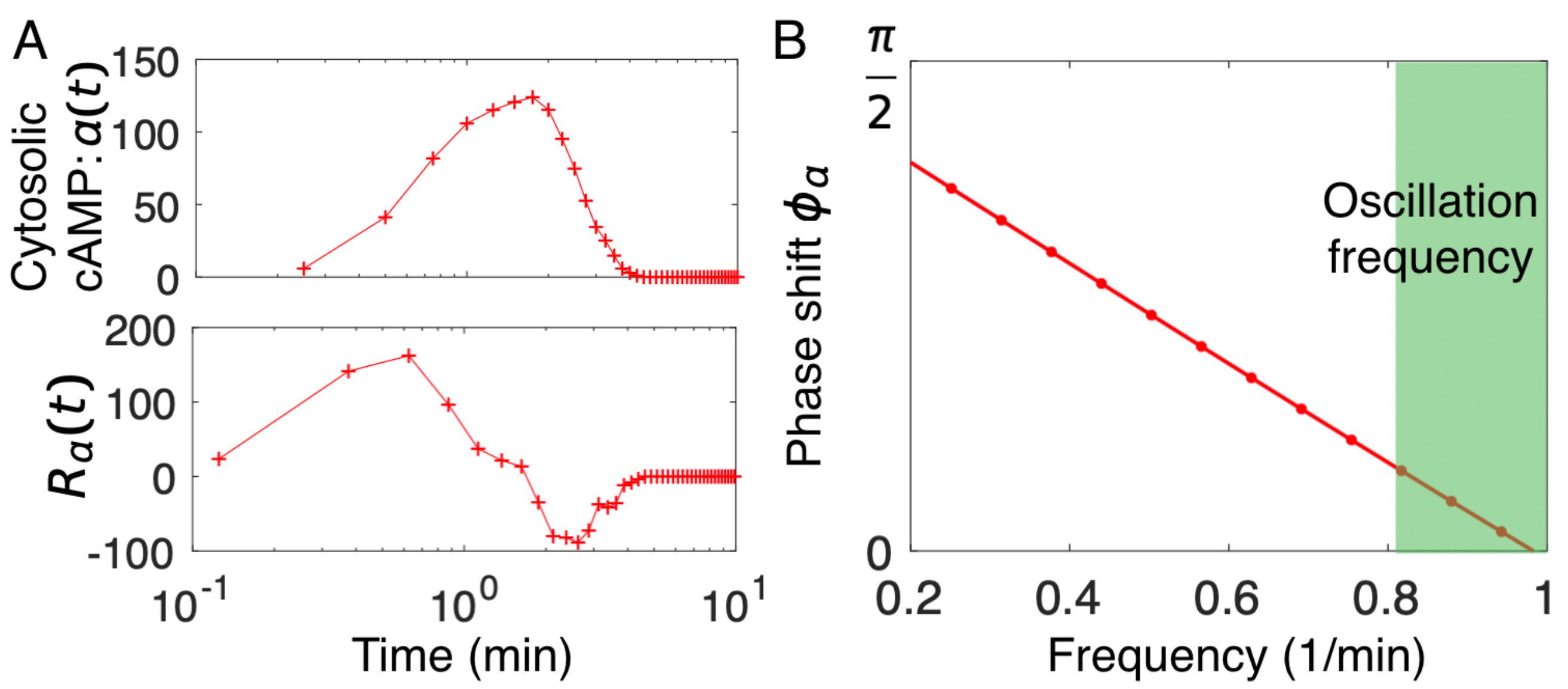}
\caption{Intracellular activity response function constructed from single-cell measurements on {\it Dictyostelium}.  
({\it A}) {\it Upper panel:} The average cytosolic cAMP level (the activity $a$, arbitrary unit) in response to a step increase of 1 nM extracellular cAMP at $t=0$  
(reproduced from Fig.~2A in Ref.~\cite{sgro2015intracellular}).  
{\it Lower panel:}  The response function $R_a(t)$ estimated from the derivative of the response data in the upper panel.  
({\it B}) The corresponding phase shift in the low frequency regime, obtained from the Fourier transform of $R_a(t)$.  
Onset oscillation frequencies in experiments span the green regime (see Fig.~2B in Ref.~\cite{gregor2010onset}). } 
\label{fig:Dicty-phase}
\end{figure}

DQS in {\it Dictyostelium} is a time-dependent phenomenon coupled to cell migration and development~\cite{schaap2011evolutionary,bretschneider2016progress}.
In the experiments reported in Refs.~\cite{gregor2010onset,sgro2015intracellular}, synchronised firing of cells starts
five hours after nutrient deprivation. The period of firing shortens from 15 - 30 min at the onset to 8 min and thereafter 6 min as
cells begin to aggregate.   Due to a property known as logarithmic sensing, receptor activity responds to the logarithm of the 
extracellular cAMP concentration $s$ instead of $s$ itself. This confers a much greater dynamical range 
in cell's adaptive response~\cite{tu2013quantitative}. Implementing logarithmic sensing in the FHN model, Sgro {\it et al.}~\cite{sgro2015intracellular} 
showed that the coupled equations are able to reproduce the accelerated pulsing when aggregation starts.  
In terms of the activity response function $R_a$, logarithmic sensing yields an amplitude $\alpha_2\sim 1/s$.
Consequently, in a steady-state situation, the signal relay efficiency $\bar{N}\sim \alpha_1N/s$ is independent of the 
cell density $N$ when the only source of $s$ is secretion by individual cells, i.e., $s\propto N$.
DQS then becomes a cell-density-independent phenomenon as noticed previously by Kamino {\it et al.} when cells 
conduct ``fold-change-detection''~\cite{kamino2017fold}.
Given that the whole developmental process takes hours to complete, intracellular signalling and gene expression may well 
undergo significant change, moving $\bar{N}$ from below to above its threshold value. 
In particular, the negative feedback time $\tau_y$, which is much longer than the activation time $\tau_a$ ($\simeq 1$ min)
of the enzyme adenylyl cyclase ACA that produces cytosolic cAMP upon extracellular cAMP stimulation,
may be affected by multiple intracellular regulatory mechanisms~\cite{devreotes2017excitable,janetopoulos2001receptor}.
These fascinating topics may have wide-ranging applications, including the working of the segmentation clock in the 
presomitic mesoderm~\cite{hubaud2017excitable}.  

Our third example, the glycolytic oscillation in yeast cell suspensions, is also an open problem. 
Simulation studies of a detailed model of yeast glycolysis~\cite{du2012steady} yielded a relatively simple phase diagram 
shown in Fig.~\ref{fig:glycolysis}{\it B}, with the extracellular glucose and acetaldehyde concentrations as control parameters.
As reported previously~\cite{du2012steady}, cells in the white regime oscillate spontaneously in a constant extracellular 
environment, driven by an instability associated with the negative feedback in ATP autocatalysis. 
In the neighbourhood of this regime, we found that the ATP concentration adapts to the extracellular environment, 
in particular to a sudden shift in acetaldehyde concentration that affects directly the intracellular NAD/NADH ratio.
The adaptation error increases as one moves away from the oscillatory regime. 
These dynamical features are captured by a 3-variable model of ATP autocatalysis we proposed to approximate the 
low-dimensional attractor of the full model at high extracellular glucose concentrations. (Technically, the reduced model 
represents a modified version of the full model where the glyoxylate shunt is turned off. The latter affects the coupling of 
ACE to NAD/NADH and more importantly ATP consumption.)
We then considered a scenario for fast equilibration of intracellular 
and extracellular ACE concentrations and comparable degradation/dilution rates in and outside the cell.
Collective oscillations were observed that span both adaptive and oscillatory regimes of a single cell. 
In this respect, our study unifies the two competing scenarios regarding
the origin of glycolytic oscillations, i.e., autonomous or through mutual stimulation. 
Cell-to-cell variability can change the size of each regime on the single-cell phase diagram.
Furthermore, the characteristic frequency $\omega^\ast$ of each cell could have a significant spread in both the
adaptive and oscillatory regimes~\cite{gustavsson2015entrainment}. 
Consequently, the range of cell densities where collective oscillations take place could be much reduced.
More work is needed to see whether the two scenarios could be separated in a heterogeneous population of yeast cells
by tuning the degradation/dilution rate of the extracellular ACE.

These model studies helped to refine and resolve various quantitative issues in the induction of collective oscillations in 
well-studied systems, and at the same time inspire novel applications built around adaptation-driven signal relay.
One promising direction to follow is the development of artificial oscillatory systems with techniques from 
synthetic biology~\cite{danino2010synchronized,liu2011sequential,potvin2016synchronous,Lim2018synthetic}. 
In analogy with the hair cell/glass fibre setup, one may think of tricking a quorum-sensing cell to oscillate
by confining it to a volume small enough to enable positive signal relay.

In statistical physics, the response function formalism is widely used to analyse system level response to environmental perturbations, but its application to collective behaviour in biological systems is still limited. 
Our examples show that cell models with different levels of biological detail,  out of either necessity or convenience, 
could yield qualitatively or even quantitatively similar response curves with respect to, say the production of a particular 
chemical used in cell-to-cell communication, which is reassuring. As these curves are increasingly accessible from experiments, 
 their direct use for analysis and hypothesis building is highly desirable.
With respect to the link between adaptation and collective oscillations, our formulation unifies and generalises previous studies 
in at least three specific settings. The first is an abstract 3-variable model that connects fold-change detection of individual cells to the robustness of collective oscillations over a broad range of cell densities~\cite{kamino2017fold}.  In the second case, adaptation was proposed to play an important role in the collective oscillation of neuronal networks~\cite{matsuoka1985sustained}.   
Lastly,  an Ising-type model of chemoreceptor arrays in \emph{E.~coli}~\cite{Mello2004-tn}  predicts that increasing the coupling strength between adaptive receptors drives the system to collective oscillations, although in reality the chemoreceptor array manages to operate below the oscillatory regime. Despite the risk of running into an oscillatory instability, the coupling enhances sensitivity of the  
array to ligand binding.  Along this sensitivity-stability tradeoff, one may speculate that some of the reported collective oscillations under laboratory conditions could  actually arise from over perfection of adaptive/homeostatic response in the 
natural environment, a hypothesis that invites further experimental testing.\\
}

\noindent
{\large \textbf{Methods}}\\
Extended materials and methods are presented in Supplementary Information. \\

\noindent
\textbf{An adaptive model with cubic nonlinearity.}
 The data presented in  Fig.~\ref{fig:weakly-adaptive-model} were obtained from numerical integration
of the coupled equations~\cite{lan2012energy,shouwen2015adaptation}: 
$\tau_a\dot{a} =-a- c_3a^3+y+\alpha_2 s+\eta_a,$ and $\tau_y\dot{y} =- a-\epsilon y+\eta_y.$ 
Here $y$ is a memory node that implements negative feedback control on $a$, 
$\epsilon$ sets the adaptation error,   and $\tau_a$ and $\tau_y$ are the intrinsic timescales for the dynamics of $a$ and $y$, respectively.  $\eta_a$ and $\eta_y$ are gaussian white noise with zero mean 
and correlations:  $\langle \eta_a(t)\eta_a(\tau)\rangle=2T \tau_a \delta(t-\tau)$ and $\langle \eta_y(t)\eta_y(\tau)\rangle=2T \tau_y \delta(t-\tau)$,  where $\delta(t)$ is the Dirac delta function.  
The cubic nonlinearity ($c_3a^3$) is needed to limit cellular activity to a finite strength.   
For simplicity, we choose $\alpha_2=1$ so that the response function defined by
$\tilde{R}_a(\omega)=\langle \tilde{a}(\omega)\rangle/\tilde{s}(\omega)$ can be compared with its equilibrium counterpart that
satisfies the FDT $\tilde{R}_a''=\omega\tilde{C}_a(\omega)/(2T)$, with $\tilde{R}_a''$ denoting the imaginary component
of $\tilde{R}_a$.  Data in Fig.~\ref{fig:weakly-adaptive-model} were obtained by coupling cells via Eq.~(\ref{eq:general-adaptation-4}) with $F(s)=-Ks$  and $\xi=0$. \\

\noindent
\textbf{Solution of the phase-matching condition Eq.~(\ref{eq:self-consistency-phase}) under an adaptive response.} 
We have shown in the Main Text that adaptive intracellular observables exhibit a phase-leading response in a certain frequency interval.  
For a given adaptive observable $a$, the phase lead $\phi_a(\omega)$ spans a continuous range from 0 to a maximum value 
$\phi_{a}^{max}$ ($<\pi$). Meanwhile, the phase delay $\phi_s=-\tan^{-1}(\omega\tau_s)$  varies continuously from 0 to $-\pi/2$ [Eq.~(\ref{eq:phi_s})].    Since $\tau_s$ controls how fast $\phi_s(\omega)$ decreases
from 0 to $-\pi/2$ as $\omega$ increases, intersection of $-\phi_s(\omega)$ with $\phi_a(\omega)$ can always be found
by tuning $\tau_s$. In particular, when $\tau_s\rightarrow 0$, a solution is found at the high frequency end of the
active frequency interval where $\phi_a(\omega)=0$. From this discussion, we see that the onset frequency $\omega_o$ of oscillations
is mostly determined by the intracellular dynamics, i.e., $\phi_a(\omega)$, but the medium can have a weak effect on
$\omega_o$ when its relaxation time is comparable to that of the intracellular dynamics.\\

\noindent
\textbf{Coupled FitzHugh-Nagumo model.}  A single FHN circuit takes the form,
$\tau_a \dot{a}_j=a_j-a_j^3/3-y_j+ \alpha_2 s+ \eta_{a_j}$, $\tau_y \dot{y}_j= a_j-\epsilon y_j+a_0+\eta_{y_j}.$
The positive sign of the first term in the equation for $a_j$ gives rise to excitability.  In the absence of the stimulus $s$, each cell assumes the ``resting state'' with a mean 
activity $a_{rs}\equiv \langle a_j(t)\rangle$.  For small values of $\epsilon$, the resting state activity $a_{rs}\simeq -a_0$ is nearly constant under 
a slow-varying $s(t)$.  FHN circuits are coupled together through a signal field whose dynamics is described by,
$\tau_s \dot{s}=-s+\alpha_1 \sum_j^N  (a_j-a_{rs})$.   The cubic nonlinear term $s^3$ is introduced to suppress discontinuous jump of the oscillation amplitude at the onset.  The parameters used in generating Fig.~\ref{fig:FHN_model} are:  $\alpha_2=1$, $N=1000$, $\epsilon=0.1$,  $T=0.1$,  $\tau_a=1$,  $\tau_y=5$,  $a_0=1.5$,   and $\tau_s=1$. $\alpha_1=\bar{N}/(N\alpha_2)$ is determined by the control parameter $\bar{N}$.   \\

\noindent
{\large \textbf{Data availability}}\\
The data that support the findings of this study are available from S.-W. Wang on request.  They can also be generated from the provided code. \\

\noindent
{\large \textbf{Code availability}}\\
The code that support the findings of this study are available at https://github.com/ascendancy09/Collective-oscillations.




\noindent\\
{\large \textbf{Acknowledgements}}
\noindent\\
The authors thank Allon Klein and Kyogo Kawaguchi for helpful discussions and suggestions on the manuscript.  
The work is supported in part by the NSFC under Grant Nos. U1430237, 11635002 and U1530401, 
and by the Research Grants Council of the Hong Kong Special Administrative Region (HKSAR) under Grant No. 12301514 and C2014-15G. 

\noindent\\
{\large \textbf{Author contributions}}\\
S.-W. W. and L.-H. T. designed research, performed research, analysed data, and wrote the paper. 

\noindent\\
{\large \textbf{Additional Information}}
\noindent\\
See Supplementary Information.

\onecolumngrid
\def\theequation{S\arabic{equation}}
\setcounter{equation}{0}


\def\thefigure{S\arabic{figure}}
\setcounter{figure}{0}
\newcommand{\wh}[1]{\widehat{#1}}
\newpage

\vskip 1cm
\centerline{
 \large{\emph{Supplementary Information for}}}\vskip 1 em
\centerline{\textbf{\large{Emergence of collective oscillations in adaptive cells}
}}
\centerline{}\vskip 0.5cm
\centerline{Shou-Wen Wang$^\ast$, Lei-Han Tang$^\ast$}
\centerline{$^\ast$Corresponding authors; Email: shouwen$\_$wang@hms.harvard.edu (S.-W. W.), lhtang@csrc.ac.cn (L.-H. T.)}
\vskip 1cm
%
%
%



\tableofcontents

\section{Introduction} 

This supplementary contains derivations of various theoretical results stated in the Main Text, 
as well as exploration of a number of model systems. In view of the recent experimental studies on dynamical quorum sensing,
we shall mainly focus on auto-induced collective oscillations of cells mediated by a chemical signal.
However, to the extent that the underlying microscopic processes afford a thermodynamic description,
our approach also applies in other physical contexts, e.g. mechanical or electrical signalling.
We are particularly interested in exploring the nonequilibrium aspects of ``activity'' dynamics in a living cell.
Concepts and tools from the recently developed stochastic thermodynamics~\cite{sekimoto2010stochastic-S} 
are used to map out the pattern of energy flow, complementing the descriptive modelling based
on rate equations.
  
The material is organised as follows. Section~\ref{sect:non-equi-adaptive}  contains various mathematical results announced in 
the Main Text regarding the nonequilibrium response of an adaptive circuit and the associated energy flow.
In Section~\ref{sect:self-consistency-scheme}, we present a self-consistenct scheme to predict 
the onset of auto-induced collective oscillations and its subsequent growth.
The response functions that appear in the discussion can in principle be measured directly in experiments.
In Section~\ref{sect:glycolysis}, we re-analyse glycolytic oscillations in
yeast cell suspensions from the perspective of linear response. The ATP negative feedback loop through the enzyme PFK 
is shown to be responsible for both adaptive and oscillatory behaviour of a single cell.
The resulting phase diagram is used to construct a low dimensional model that reproduces the main dynamical features of the 
full model when the ATP feedback loop is strongly coupled to the intracellular concentration of acetaldehyde, a small
molecule that diffuses fast through the cell membrane and hence can be used as a metabolic signal.

\section{Nonequilibrium thermodynamics of adaptive response}
 \label{sect:non-equi-adaptive} 

\subsection{Phase-leading response of an adaptive variable} 

Consider the temporal variation $a_t$ of an intracellular observable $a$ induced by a sinusoidal signal $s_t$
from the environment at frequency $\omega$. The amplitude of $s_t$ is assumed to be small, so that it can be
treated as a perturbation. The observable $a$ is either directly or indirectly coupled to the signal $s$. 
On the scale of a single cell, both $a_t$ and $s_t$ may contain stochastic components.
In the following, we shall examine the noise averaged response of $a$ to the deterministic part of $s$, i.e., the signal.
As usual, we use $\langle\cdot\rangle$ to denote the noise average. 
The following convention on forward and inverse Fourier transforms is adopted, 
\begin{equation}
\tilde{f}(\omega)=\int_{-\infty}^{\infty} f(t)\exp(i\omega t)dt,\quad 
f(t)=\int_{-\infty}^{\infty} \tilde{f}(\omega)\exp(-i\omega t) \frac{d\omega}{2\pi}.
\end{equation} 

In a steady-state, the ratio of the Fourier amplitudes $\langle\tilde{a}(\omega)\rangle$
and $\langle\tilde{s}(\omega)\rangle$ defines the response function 
$\tilde{R}_a(\omega)\equiv\langle \tilde{a}(\omega)\rangle/\langle \tilde{s}(\omega)\rangle$, 
which can be separated into its real $\tilde{R}_a'$ and imaginary $\tilde{R}_a''$ parts.
(For a cellular variable $a$ that follows stochastic dynamics, the ensemble averaged response is considered.)
The well-known Kramers-Kr{\" o}nig relation from causality requirement on the response
function states~\cite{Sethna2006entropy-S}: 
\begin{equation}
\tilde{R}_a'(\omega)=\frac{2}{\pi} \int_0^\infty \tilde{R}_a''(\omega_1)\frac{\omega_1}{\omega_1^2-\omega^2}d\omega_1.
\label{eq:kramers}
\end{equation}
In the case of a perfectly adapting $a$, the response vanishes under a sufficiently slow stimulus, i.e., 
$\lim_{\omega\rightarrow 0}\tilde{R}_a(\omega)=0$.  Equation (\ref{eq:kramers}) then requires,
\begin{equation}
\label{eq:adaptation_cond}
\int_0^\infty \tilde{R}_a''(\omega_1)\omega_1^{-1}d\omega_1=0.  
\end{equation}
Consequently, $\tilde{R}_a''(\omega)$ must change sign at least once along the frequency axis. 
Let $\phi_a=-\arg(\tilde{R}_a)$ be the phase of $-\tilde{R}_a(\omega)$, with the minus sign introduced by convention.  
Positive and negative values of $\tilde{R}_a''$ thus translate to phase-lag ($-\pi<\phi_a<0$) 
and phase-lead ($0<\phi_a<\pi$) of $a_t$ over $s_t$, respectively.  
By virtue of continuity, the sign change of $\tilde{R}_a''(\omega)$ is also expected in the partially adaptive case,
provided the adaptation error $\epsilon\simeq \tilde{R}_a'(0)$ is sufficiently small.  

%

\subsection{Energy outflow from an adaptive channel}

Auto-induced collective oscillations in a dissipative medium require an energy source.  
Below, we show that an active cell is able to output energy to a fluctuating $s$ in the presence of an adaptive channel.
The power of the output depends on the strength of the coupling as well as the amplitude and frequency 
of the fluctuating signal.

Consider a slightly more general form of Eq.~(1) in the Main Text where the contribution from cell $j$ to the total thermodynamic
force on $s$ is given by $O(a_j)$, which in general is nonlinear in $a_j$. 
The work done on $s$ by the cell in a time interval $(0, L)$ can then be written as
\begin{equation}
W_j=\int_0^L  O_t\dot{s}_tdt,
\label{eq:Ws_j}
\end{equation}
where $O_t\equiv O(a_j(t))$ and $s_t$ are both fluctuating quantities in general. 
We now consider a sinusoidal signal $s_t=s_0+\Delta s \cos(\omega t)$ 
with a small amplitude $\Delta s$. To the first order in $\Delta s$, we have
\begin{equation}
O_t \simeq  O^{(0)}_t + \Delta s|\tilde{R}_O(\omega)|\cos\bigl(\omega t +\phi_O(\omega)\bigr) .
 \label{eq:O_t-response}
\end{equation}
Here $O^{(0)}_t$ denotes the stochastic trajectory of $O$ in the absence of the sinusoidal signal.
As usual, the linear response function $R_O$ in the steady-state (ss) to a weak time-varying signal $s_t$ is introduced through 
\begin{equation}
\langle O_t\rangle\simeq \langle O\rangle_{ss}+ \int_{-\infty}^t R_O(t-\tau)s_\tau d\tau,
\end{equation}
where $\langle\cdot\rangle$ denotes average over noise. The phase angle $\phi_O(\omega)\equiv -\arg\tilde{R}_O(\omega)$.
Substituting expression (\ref{eq:O_t-response}) into Eq. (\ref{eq:Ws_j}) and taking the limit $L\rightarrow\infty$,
we obtain the time-averaged output power from the cell through this channel (omitting the subscript $j$),
\begin{equation}
\overline{\dot{W}}=\overline{O_t\dot{s}_t} + o\bigl((\Delta s)^2\bigr)
=\frac{1}{2}\omega|\tilde{R}_O(\omega)|\sin\phi_O(\omega)(\Delta s)^2+o\bigl((\Delta s)^2\bigr).
\label{eq:general-W-O}
\end{equation}
Here the overline bar indicates averaging over time, and $o\bigl((\Delta s)^2\bigr)$ denotes terms 
higher than second order in $\Delta s$.

Given the relation $O_t=O(a_t)$, adaptation of the cellular variable $a$ to a slow-varying $s$ also 
implies the adaption of $O$ to the signal.  The causality condition (\ref{eq:kramers}) applied to
$O_t$ then requires $\tilde{R}_{O}''(\omega)\equiv-|\tilde{R}_O(\omega)|\sin\phi_O(\omega)<0$ in a certain frequency range.  Consequently, Eq. (\ref{eq:general-W-O}) predicts energy outflow from the channel under a periodic stimulation at
these frequencies.  

The discussion leading to Eq. (\ref{eq:general-W-O}) in the previous section can be easily extended to the energy outflow 
under an arbitrary signal variation $s_t$ with a power spectrum $\tilde{C}_s(\omega)$,
\begin{equation}
\label{eq:general-W-O1}
\overline{\dot{W}}=-\int d\omega \omega\tilde{R}_O''(\omega)\tilde{C}_s(\omega)+o(\Delta s^2),
\end{equation}
where $\Delta s$ sets the overall amplitude of signal variation. 
If the cell were in thermal equilibrium, $a$ would respond passively to a time-varying signal with a phase-lag and dissipate the
energy inflow generated by the stimulation. An adaptive cell, on the other hand, is able to output energy
in the form of work when stimulated in the right frequency range. This form of energy outflow 
is different from the heat dissipation arising from keeping the system out of equilibrium as studied in 
Refs.~\cite{lan2012energy-S,Pablo2015Adaptation-S,shouwen2015adaptation-S}.

\subsection{The Fluctuation-Dissipation Theorem}

The fluctuation-dissipation theorem (FDT) is generally presented as an identity between the response function
of a chosen variable to an external perturbation and the correlation function of the variable in question with 
the one that is conjugate to the perturbation\cite{kubo1966fluctuation-S}.
For Markov systems which are of interest here, FDT holds when the
detailed balance condition on the state-space transition rates is fulfilled. We refer the reader to 
Refs.~\cite{diezemann2005fluctuation-S,Wang2016entropy-S} for a detailed discussion, including more rigorous
definitions of various quantities of interest.

Assuming that the signal $s$ affects the cell through coupling to a conjugate variable $O$ which is
proportional to the variable $a$ of interest, i.e., $O=c_0 a$ with $c_0$ a proportionality constant.  
In this case, FDT states that
\begin{equation}
\label{eq:FDT}
\tilde{R}_O''(\omega)=\frac{\omega \tilde{C}_O(\omega)}{2T}>0.
\end{equation} 
Here, $\tilde{C}_O(\omega)=c_0^2 \langle |\tilde{a}(\omega)|^2\rangle$ is the power spectrum of $O_t$ which is always positive.   
Equation (\ref{eq:FDT}) contradicts (\ref{eq:adaptation_cond}), re-affirming that receptor adaptation cannot be realised
without the presence of active processes inside the cell. 

In Ref.~\cite{ma2009defining-S}, adaptation through a 3-node incoherent feed-forward motif was considered.
It was later shown that the topology even supports adaptation in an equilibrium setting~\cite{de2013unraveling-S}.
The main difference between these models and the adaptive receptor model  in the Main Text (Fig.~3 and Methods)  is that, 
in the former, $s$ not only couples to $a$ directly, but also to other intracellular variables. 
The conjugate variable $O$ is then a combination of $a$ and other intracellular variables. 
We leave a detailed investigation of this issue to future work.

\section{A self-consistent scheme for frequency selection and oscillation amplitude determination}
\label{sect:self-consistency-scheme}

The thermodynamic analysis in the preceding section suggests the possibility of a positive feedback loop
formed by a periodic signal and adaptive cells under generic conditions. Collective oscillations emerge when
signal amplification by active cells overtakes signal dissipation in the passive medium. In this section, we examine
this process in further detail and derive equations that can be used to determine the frequency and amplitude of
auto-induced oscillations when the instability takes place. For simplicity, we shall consider a situation where diffusion of
the signalling molecules in the medium is very fast so that spatial variations of $s$ is suppressed. Consequently, 
the notion of a well-defined transition to the oscillating state can be introduced. 

\subsection{The phase matching condition and threshold cell density}


Given that individual cells couple to each other only through the signal field $s$, a self-consistency procedure similar to the solution
of mean-field models in statistical physics can be employed. In this case, the linear equations governing an eigenmode 
with eigenvalue $\lambda$ can be divided into subgroups associated with individual cells. 
The internal variables of a given cell appear in one and only one of 
the subgroups. Solution of the subset of equations for cell $j$ yields the cell activity $\langle \tilde{a}_j\rangle=\tilde{R}_{a,j}(i\lambda)\langle \tilde{s}\rangle$.
The function $\tilde{R}_{a,j}(i\lambda)$ is the same function introduced in the preceding
section to describe the linear response of $a_j$ to a sinusoidal perturbation at frequency $\omega=i\lambda$.
Likewise, a response function $\tilde{R}_s(\omega)$ from the linearised relaxational dynamics of $s$ can be obtained,
treating contributions from cells as source terms, as in Eq. (1) of the Main Text. 
Combining the two steps, we arrive at the following eigenvalue equation
for $\lambda$,
\begin{equation}
\tilde{R}_s(i\lambda)\sum_{j=1}^N\alpha_1\tilde{R}_{a,j}(i\lambda)=1.
\label{eq:eigenvalue_eq}
\end{equation}
When a particular eigenvalue crosses the imaginary axis, its real part vanishes, while its imaginary part $\omega_o$ (the onset frequency) satisfies,
\begin{equation}
\alpha_1N_o\tilde{R}_s(\omega_o)\tilde{R}_{\bar{a}}(\omega_o)=1.
\label{eq:eigenfrequency_onset}
\end{equation}
Here $\tilde{R}_{\bar{a}}(\omega)\equiv N^{-1}\sum_{j=1}^N\tilde{R}_{a,j}(\omega)$ is the averaged single-cell response
function.

Equation (\ref{eq:eigenfrequency_onset}) can be written separately for the phase shift $\phi=-\arg{\tilde{R}}$ and
amplitude $|\tilde{R}|$ of the response functions. For $\alpha_1>0$, we have,
\begin{subequations}\label{eq:oscillation-prediction}
\begin{eqnarray}
\label{eq:phase-matching}
\phi_{\bar{a}}(\omega_o)&=& -\phi_s(\omega_o),\\
\label{eq:amplitude}
N_o&=&\frac{1}{|\alpha_1 \tilde{R}_{s}(\omega_o)||\tilde{R}_{\bar{a}}(\omega_o)|}.
\end{eqnarray}
\end{subequations}
Eq. (\ref{eq:phase-matching}) determines the frequency $\omega_o$ at the onset of collective oscillations,
while Eq. (\ref{eq:amplitude})  gives the threshold cell density $N_o$. 
As we mentioned in the Main Text, when the signal is passive, phase lead by the cell is required for Eq. (\ref{eq:phase-matching})
to be fulfilled. The explicit relation presented here complements the energy argument based on Eq. (\ref{eq:general-W-O}),
with the activity-generated thermodynamic force $O_t$ being proportional to  $\alpha_1 a_j$.

As it stands, the cell density $N$ does not appear explicitly in the phase-matching condition
(\ref{eq:phase-matching}). Therefore the frequency of collective oscillations can be estimated from separate measurements of
the single-cell response and the medium response. In reality, it is conceivable that properties of the medium
are affected by the presence of cells, e.g., the concentration of the signalling molecules secreted. 
Consequently, both $\tilde{R}_{s}(\omega)$ and $\tilde{R}_{\bar{a}}(\omega)$ may have certain weak dependence on $N$.

\subsection{The amplitude equations and frequency shift}

Beyond the initial instability, nonlinear effects need to be treated explicitly to determine the
amplitude and frequency of oscillations. Assuming a periodic state, the signal strength $s(t)$ can be
expressed as a Fourier series that includes the first harmonic as well as higher order harmonics produced by nonlinearities
in the system dynamics. Likewise, the noise-averaged cellular activity $\langle a_j(t)\rangle$ can also be expressed as a 
Fourier series in $t$ with the same basic frequency. For weak noise, the trajectory of the system falls on a well-defined
limit cycle whose mean radius $r$ sets the overall amplitude of oscillations, while the amplitude of the $n$th order harmonic
scales as $r^n$. This structure allows for a systematic determination of the amplitudes using perturbation theory. 
Below, we illustrate the procedure in the case of cubic nonlinearities in both the dynamics 
for $s$ and the dynamics for $a$, and comment on similarities and differences in more general situations. 
When the cell's activity is noisy, more sophisticated schemes based on the probability distribution function
of the cellular state need to be introduced (see, e.g. Ref.~\cite{diezemann2012NL_response-S}).

Let us consider a noiseless version of the adaptive dynamics defined in the Main Text (Fig.~3 and Methods), together with
a modified version of Eq. (1) that includes a cubic nonlinearity,
\begin{subequations}
\label{eq:nonlinear-a-s-oscillation}
\begin{eqnarray}
\tau_a\dot{a}_j&=&-(a_j-y_j)-c_3a_j^3+\alpha_2 s\\
\tau_y \dot{y}_j&=&- (a_j+\epsilon y_j)\\
\tau_s \dot{s}&=&- s-d_3 s^3+\alpha_1 \sum_{j=1}^N  a_j.
\end{eqnarray}
\end{subequations}
 Here $\tau_s=\gamma_s/K_s$ gives the relaxation timescale for the signal.  We also set $\alpha_1\to K_s\alpha_1$ for notational simplicity.  The two coefficients
$c_3$ and $d_3$ set the strengths of nonlinearities in the cellular and signal dynamics, respectively. 
The model has the inversion symmetry $s\rightarrow -s$ and $(a_j,y_j)\rightarrow (-a_j,-y_j)$, all $j$.
Furthermore, if we redefine the sign of $s$ and at the same time change the sign of $\alpha_2$ and $\alpha_1$,
the equations remain the same. 

We now seek a periodic solution to Eqs. (\ref{eq:nonlinear-a-s-oscillation}) in Fourier form, 
\begin{subequations}
\label{eq:signal-Fourier-series}
\begin{eqnarray}
s(t)&=& B\cos(\omega t) + \sum_{n=2}^\infty B^{(n)}\cos(n\omega t +\phi_s^{(n)}),\\
a_j(t) &=& A_j\cos(\omega t +\phi_{a,j}) + \sum_{n=2}^\infty A_j^{(n)}\cos(n\omega t +\phi_{a,j}^{(n)}), 
\qquad j= 1,\ldots, N,\\
y_j(t) &=& C_j\cos(\omega t +\phi_{y,j}) + \sum_{n=2}^\infty C_j^{(n)}\cos(n\omega t +\phi_{y,j}^{(n)}), 
\qquad j= 1,\ldots, N.
\end{eqnarray}
\end{subequations}
The amplitudes and phase shifts, all assumed to be real, satisfy a set of equations which can be derived by
substituting Eqs.~(\ref{eq:signal-Fourier-series}) into Eqs.~(\ref{eq:nonlinear-a-s-oscillation}), 
and grouping terms according to the order of the harmonic.

Starting from the first harmonic in the expressions (\ref{eq:signal-Fourier-series}), the cubic terms in 
Eqs.~(\ref{eq:nonlinear-a-s-oscillation}a) and (\ref{eq:nonlinear-a-s-oscillation}c) generate the first and third order harmonics
according to the identity $(\cos\phi)^3=(3\cos\phi+\cos3 \phi)/4$. Hence terms such as $A_j^3$ and $B^3$ are present in the
equations for the first harmonic. On the other hand, the cubic nonlinearities do not generate even order harmonics if they are not
included in the series initially. Hence, up to the third order in the amplitudes, the equations for the coefficients of
the first harmonic take the form,
\begin{subequations}
\label{eq:cubic-expansion}
\begin{eqnarray}
-i\omega \tau_a \tilde{a}_j&\simeq& - (1+{3\over 4}c_3 |\tilde{a}_j|^2)\tilde{a}_j + \tilde{y}_j +\alpha_2 \tilde{s},\\
-i\omega \tau_y \tilde{y}_j&=& - \tilde{a}_j - \epsilon\tilde{y}_j ,\\
-i\omega \tau_s \tilde{s}&\simeq& - (1+{3\over 4}d_3 |\tilde{s}|^2)\tilde{s}
+\alpha_1 \sum_j \tilde{a}_j.
\end{eqnarray}
\end{subequations}
Here we have introduced the short-hand notations $\tilde{s}=B,\tilde{a}_j=A_j\exp(-i\phi_{a,j})$, 
and $\tilde{y}_j=C_j\exp(-i\phi_{y,j})$. 

To gain an intuitive understanding of the oscillatory solution as the cell density increases
beyond the threshold $N_o$, we first eliminate the intracellular variable $\tilde{y}_j$ in Eqs. (\ref{eq:cubic-expansion}a)
and (\ref{eq:cubic-expansion}b) to obtain,
\begin{equation}
\label{eq:amplitude-a}
\tilde{a}_j=\tilde{R}_{a,j}^+(\omega)\tilde{s},
\end{equation}
where 
\begin{equation}\label{eq:Ra-nonlinear}
\tilde{R}_{a,j}^+(\omega)\equiv {\tilde{a}_j(\omega)\over \tilde{s}(\omega)}\simeq 
\frac{\alpha_2}{1+3c_3 |\tilde{a}_j|^2/4-i\omega\tau_a-1/(i \omega \tau_y-\epsilon)}
\end{equation}
is a ``nonlinear response function'' which expresses the ratio of the complex amplitudes of the first harmonic on
the limit cycle.
Similarly, Eq. (\ref{eq:cubic-expansion}c) can be rewritten as
\begin{equation}
\label{eq:amplitude-s}
\tilde{s}=\tilde{R}_{s}^+(\omega)\sum_{j=1}^N\alpha_1\tilde{a}_j,
\end{equation}
where 
\begin{equation}
 \tilde{R}_s^+(\omega)\simeq \frac{1}{1+3d_3 |\tilde{s}|^2/4-i\omega\tau_s}
 \label{eq:Rs*}
 \end{equation}
is a ``nonlinear response function'' of $s$ on the limit cycle. It is easy to see that 
$\tilde{R}_a^+(\omega)$ and $\tilde{R}_s^+(\omega)$ reduce to their respective linear counterparts $\tilde{R}_{a,j}(\omega)$ and
$\tilde{R}_s(\omega)$ when the oscillation amplitudes vanish.

We now combine Eqs. (\ref{eq:amplitude-a}) and (\ref{eq:amplitude-s}) to obtain the self-consistency condition,
\begin{equation}
\alpha_1N\tilde{R}_s^+(\omega)\tilde{R}_{\bar{a}}^+(\omega)=1,
\label{eq:eigenfrequency_LC}
\end{equation}
which is reminiscent of Eq. (\ref{eq:eigenfrequency_onset}).
Here $\tilde{R}_{\bar{a}}^+(\omega)\equiv N^{-1}\sum_{j=1}^N\tilde{R}_{a,j}^+(\omega)$ is 
the averaged single-cell nonlinear response function. When all cells are identical, 
$\tilde{R}_{\bar{a}}^+(\omega)=\tilde{R}_a^+(\omega)$.
As before, Eq.~(\ref{eq:eigenfrequency_LC}) can be rewritten
in terms of the phase and amplitude of the nonlinear response functions,
\begin{subequations}\label{eq:oscillation-prediction-LC}
\begin{eqnarray}
\label{eq:phase-matching-LC}
\phi_{\bar{a}}^+(\omega)&=& -\phi_s^+(\omega),\\
\label{eq:amplitude-LC}
\alpha_1 N&=&\frac{1}{|\tilde{R}_{s}^+(\omega)||\tilde{R}_{\bar{a}}^+(\omega)|}.
\end{eqnarray}
\end{subequations}
Formally, Eq. (\ref{eq:phase-matching-LC}) can be used to determine the frequency shift at a finite amplitude of oscillation,
while Eq. (\ref{eq:amplitude-LC}) relates the oscillation amplitude to the cell density $N$. Since the amplitudes enter
quadratically into the nonlinear response functions, they are expected to increase as $(N-N_o)^{1/2}$ just above
the threshold cell density $N_o$, e.g., the transition is of the Hopf bifurcation type.

\begin{figure}[!h]
\centering
\includegraphics[width=12cm]{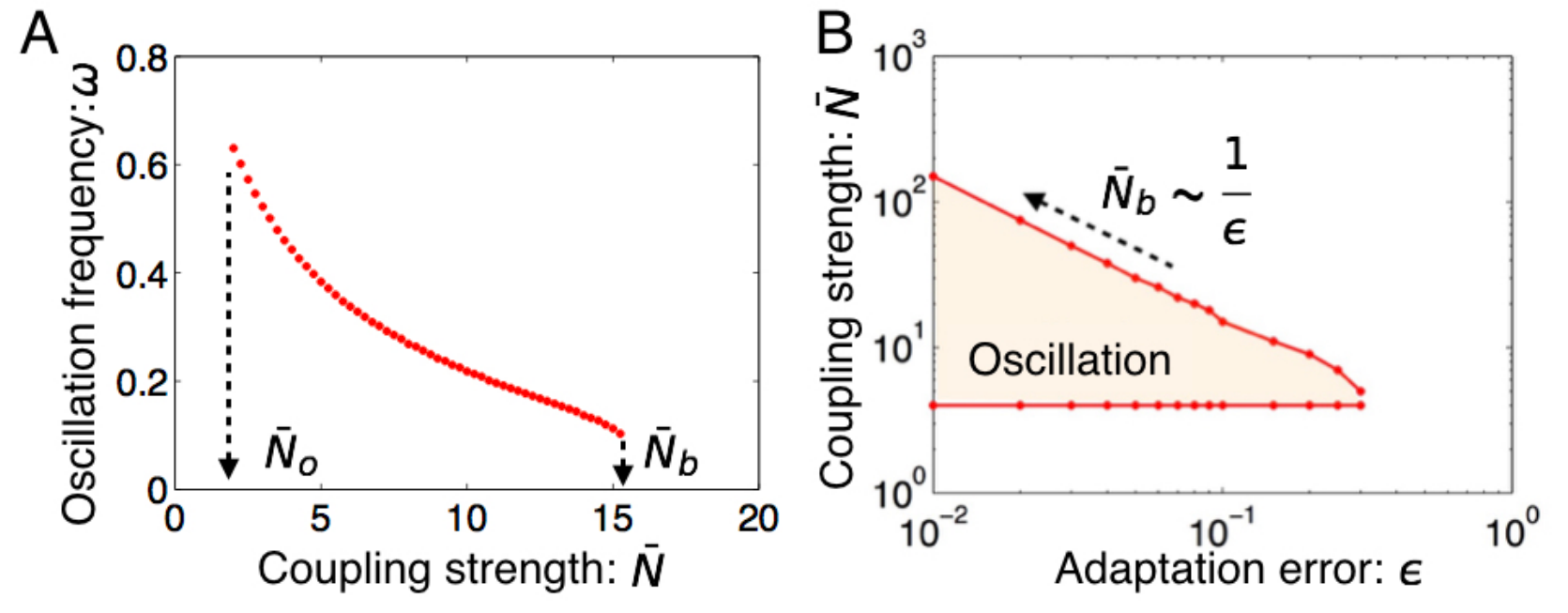}
\caption{\textbf{Collective oscillations in the model Eq.~(\ref{eq:nonlinear-a-s-oscillation})  
with nonlinear adaptation ($c_3=1$) and linear signal relaxation ($d_3=0$)}. 
({\it A}) Oscillation frequency against the effective cell density $\bar{N}=\alpha_2\alpha_1 N$.  ({\it B}) The phase diagram 
in the plane spanned by $\bar{N}$ and the adaptation error $\epsilon$.  
Other parameters are the same as in Fig.~3 of the Main Text.  }
\label{fig:adaptation_linear_signal}
\end{figure}

In the Main Text, we have considered the case $c_3=1$ and $d_3=0$. Numerically, the oscillation frequency is found to decrease 
as the coupling strength $\bar{N}\equiv\alpha_2\alpha_1 N$ increases (see also Fig.~\ref{fig:adaptation_linear_signal}{\it A}). 
This is consistent with Eq.~(\ref{eq:phase-matching-LC}) whose solution at selected oscillation amplitudes is shown 
in Fig.~3{\it F} in the Main Text. As the amplitude of the oscillations increase, $\phi_a^+(\omega)$ decreases on the low frequency side.
Consequently, the intersection point with $\phi_s^+(\omega)=\phi_s(\omega)$ shifts to lower frequencies.

Interestingly, the limit cycle associated with the oscillating state in this model shrinks to a fixed point when $\bar{N}$ 
exceeds an upper threshold value $\bar{N}_b$. The dependence of $\bar{N}_b$ on the adaptation error $\epsilon$, which
is assumed to be small, can be estimated as follows.
At a fixed point of Eqs.~(\ref{eq:nonlinear-a-s-oscillation}) at $d_3=0$, we have
$s\simeq \alpha_1 N a$ (from $\dot{s}=0$),  $y\simeq \alpha_2 s  $ (from $\dot{a}=0$),  
and $a\simeq \epsilon y$ (from $\dot{y}=0$).  Consequently, the upper threshold for oscillations has the scaling
\begin{equation}
\bar{N}_b\sim \frac{1}{\epsilon}.
\label{eq:NH-epsilon}
\end{equation}
Figure~\ref{fig:adaptation_linear_signal}{\it B} shows the numerical values for $\bar{N}_o$ and $\bar{N}_b$ 
against the adaptation error $\epsilon$ obtained in our simulations, which confirms (\ref{eq:NH-epsilon}).
The oscillating state expands over a larger range of cell densities when individual cells are more adaptive.

\begin{figure}[!h]
\centering
\includegraphics[width=12cm]{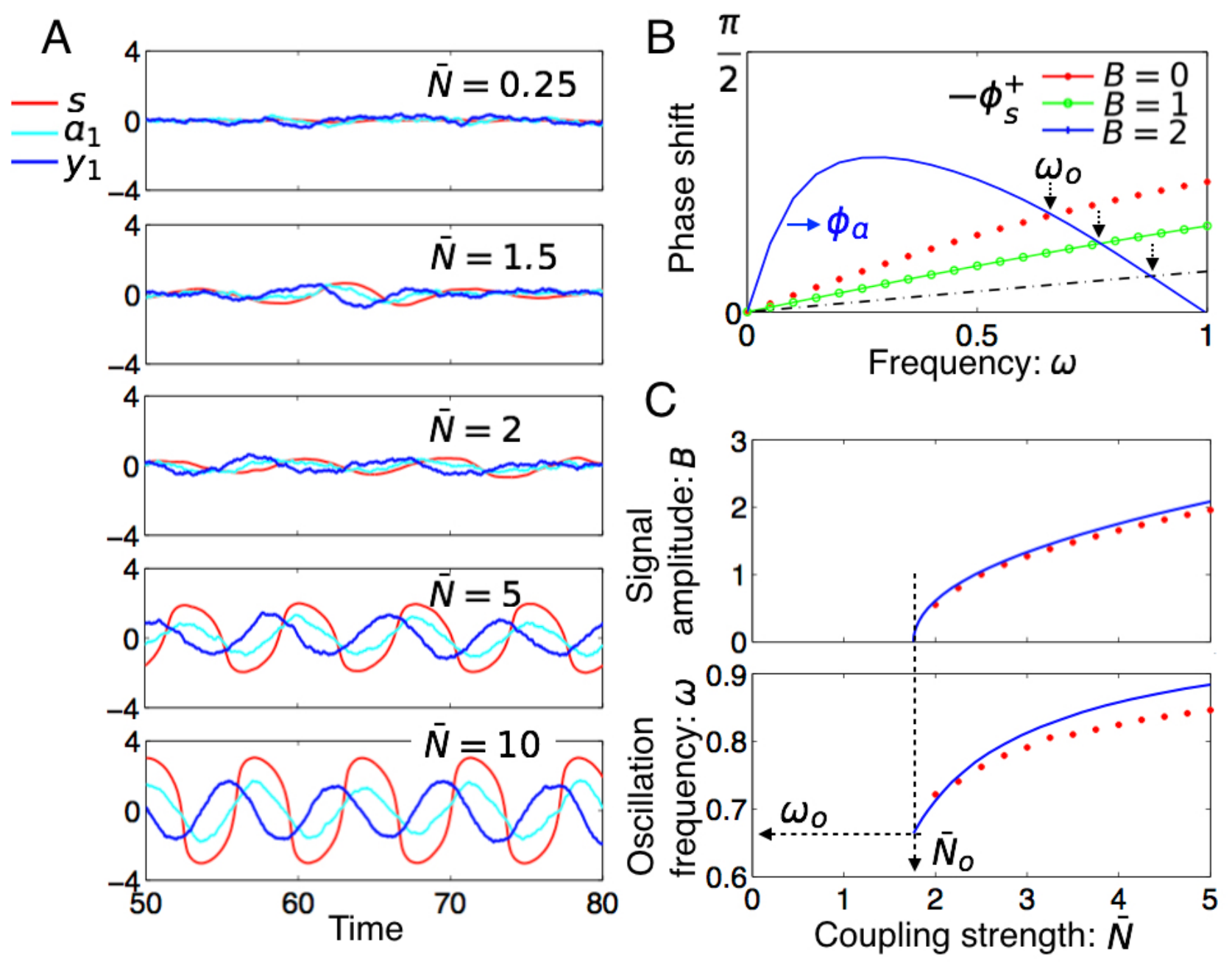}
\caption{\textbf{Collective oscillations in the model Eq.~(\ref{eq:nonlinear-a-s-oscillation})  with linear adaptation ($c_3=0$) and nonlinear signal relaxation ($d_3=1$)}. 
({\it A}) Temporal trajectories at various values of $\bar{N}$.  ({\it B}) The phase lead $\phi_a$ and lag $-\phi_s^+$ against
$\omega$ at selected oscillation amplitudes.  ({\it C}) The predicted oscillation frequency and amplitude as compared with those obtained from numerical simulations.  Other parameters are the same as in Fig.~3 of the Main Text.   }
\label{fig:adaptation_nonlinear_signal}
\end{figure}

Next, consider the case of nonlinear signal relaxation ($d_3=1)$ and linear adaptation ($c_3=0$).  
The onset of collective oscillations is similar to the previous case (Fig.~\ref{fig:adaptation_nonlinear_signal}{\it A}), 
except that oscillations speed up as the cell density increases further.  From Eq.~(\ref{eq:Rs*}), we obtain
\begin{equation}
\phi_s^+(\omega)=-\arg \tilde{R}_s^+(\omega)=-\arctan \Big[\frac{\omega}{\omega_s(1+3d_3 |\tilde{s}|^2/4)}\Big],
\end{equation}
which decreases as the oscillation amplitude increases. As shown in Fig.~\ref{fig:adaptation_nonlinear_signal}{\it C},
the intersection point shifts to the right. The predicted signal oscillation amplitude $B=|\tilde{s}|$ and frequency shift agree 
quantitatively with our numerical results (Fig.~\ref{fig:adaptation_nonlinear_signal}{\it C}).   

In the more general case when both $c_3$ and $d_3$ are nonzero, we need to first express
$\tilde{s}$ and $\tilde{a}_j$ in terms of a common variable that specifies oscillation amplitude
before applying the phase-matching condition Eq.~(\ref{eq:phase-matching-LC}).
As we see from the discussions above, depending on which of the two cubic nonlinearities is stronger, the
oscillation frequency may shift either to lower or higher values. In general, nonlinearities may also be present 
in the dynamics of other intracellular variables which need to be dealt with case by case.

When quadratic nonlinearities are present in the system dynamics, the second harmonic is generated and need to be
considered in the perturbative analysis. Consider for example the equation for $a_j$ with an extra term $c_2a_j^2$.
Following the same procedure that led to Eqs.~(\ref{eq:cubic-expansion}), we find an additional term
$c_2\tilde{a}_j^*\tilde{a}_j^{(2)}$ on the right-hand side of Eq.~(\ref{eq:cubic-expansion}a), where $\tilde{a}_j^{(2)}$
is the amplitude of the second harmonic in $a_j(t)$ (including phase). 
The amplitude equation for the second harmonic relates $\tilde{a}_j^{(2)}$ to $c_2\tilde{a}_j^2$ and $\alpha_2\tilde{s}^{(2)}$. 
Together with the equation for $\tilde{s}^{(2)}$, amplitudes of the second harmonic can be expressed as a linear
combination of terms $c_2\tilde{a}_j^2$ from different cells. The upshot of this exercise is that coefficient of the
cubic term $|\tilde{a}_j|^2\tilde{a}_j$ in Eq.~(\ref{eq:cubic-expansion}a) should contain additional contributions
proportional to $c_2^2$. The nonlinear response functions (\ref{eq:Ra-nonlinear}) and (\ref{eq:Rs*}) on the limit
cycle can still be defined in the same way, and Eq.~(\ref{eq:eigenfrequency_LC}) still holds formally.
Through $\tilde{s}^{(2)}$, terms $|\tilde{a}_k|^2$ from other cells enter the expression for $\tilde{R}_{a,j}^+(\omega)$.
Two conclusions can be drawn from this fact: i) as in the case of cubic nonlinearities, the transition is still of the Hopf 
bifurcation type; ii) $\tilde{R}_{a,j}^+(\omega)$ can no longer be determined by simply measuring the
response of a given cell to a sinusoidal stimulus at finite strength, as it is affected by the oscillation pattern of
other cells in the system due to the quadratic nonlinearity.

\subsection{Requirement on the speed of signal relaxation/clearance for the onset of collective oscillations} 
\label{sect:adaptation-error-timescale}

Experiments have indicated that sufficiently fast breakdown of the signalling molecule is needed for  DQS
in dicty~\cite{gregor2010onset} and for sustained oscillations in yeast cell suspensions~\cite{richard1994yeast-S}.
Below we derive an upper limit for the signal relaxation time $\tau_s$ to satisfy the phase matching condition
Eq.~(4a) in the Main Text. The result is inversely proportional to the adaptation error $\epsilon$ of the intracellular circuit.

\begin{figure}
\centering
\includegraphics[width=8.5cm]{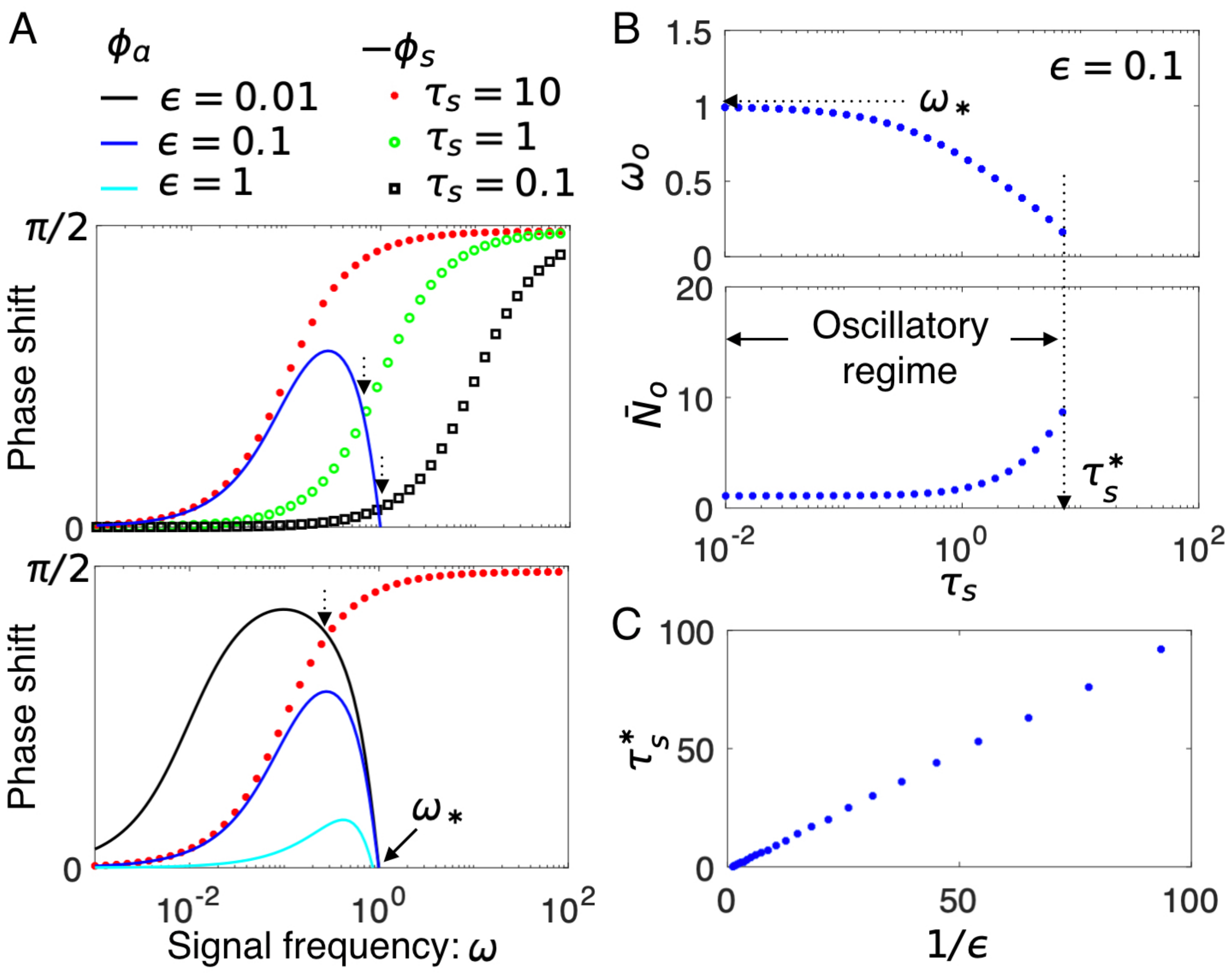}
\caption{Phase-matching at the onset of oscillations for different values of signal timescale $\tau_s$ and
adaptation error $\epsilon$.    ({\it A}) Phase shift of the cell activity ($\phi_a$) and signal response ($|\phi_s|$) at selected values of
$\tau_s$ (upper panel) and adaptation error $\epsilon$'s (lower panel).   
The onset frequency is given by the intersection of the two curves [Eq.~(\ref{eq:phase-matching})].  ({\it B}) Predicted onset frequency $\omega_o$ and onset coupling strength $\bar{N}_o$ for different values of $\tau_s$.    
Oscillations will not be found when the signal relaxation time $\tau_s>\tau_s^*(\epsilon)$.  
({\it C})  Numerical results supporting linear scaling between $\tau_s^*(\epsilon)$ and $1/\epsilon$.  
The data are obtained from the coupled adaptive circuits under the same parameters (except $\epsilon$ and $\tau_s$) as in Fig.~3 in the Main Text. 
}
\label{fig:effect_of_taus}
\end{figure}

Taking Eq.~(5) for the signal phase shift, $\phi_s=-\tan^{-1}(\omega \tau_s)$, we see that
a longer $\tau_s$ yields a larger signal delay $|\phi_s|$ at a given frequency.
This is illustrated by Fig.~\ref{fig:effect_of_taus}{\it A}, upper panel, where the horizontal frequency axis is shown on logarithmic scale.
For a given $\epsilon$, the intersection of the two phase-shift curves moves to the left, yielding a lower
onset oscillation frequency $\omega_o$ and a larger coupling strength $\bar{N}$ (Fig.~\ref{fig:effect_of_taus}{\it B}).  
When $\tau_s$ reaches beyond an upper limit $\tau_s^*(\epsilon)$, the solution disappears (Fig.~\ref{fig:effect_of_taus}{\it B}).  
Interestingly, a reduction of $\epsilon$ in Eq.~(9) increases $\phi_a(\omega)$ on the low frequency side, and rescues the solution (Fig.~\ref{fig:effect_of_taus}{\it A}, lower panel).   

The observed inverse dependence of $\tau_s^*$ on $\epsilon$ in Fig.~\ref{fig:effect_of_taus}{\it C} can be justified 
from the behaviour of the two phase-shift functions at low frequencies. 
Close to $\omega=0$, Eq.~(8) in the Main Text yields $\tilde{R}_a'(\omega)\propto \epsilon$ 
while $\tilde{R}_a''(\omega)\propto \omega$, as $\tilde{R}_a''$ must be an odd function of $\omega$. 
This is confirmed by expanding Eq.~(9) in the Main Text at $\omega=0$. 
Consequently, $\phi_a(\omega)\simeq -\tilde{R}_a''/\tilde{R}_a'\propto \omega/\epsilon$.
On the other hand, $\phi_s(\omega)\approx -\omega\tau_s$ in this regime.
The two curves has an intersection at low frequencies provided
\begin{equation}
\tau_s<\tau_s^*\propto 1/\epsilon.
\label{eq:epsilon-tau_s}
\end{equation}
Using the explicit expressions Eqs. (3) and (9) in the Main Text, we obtain $\tau_s^*\simeq\tau_y/\epsilon$ for small $\epsilon$.
The critical cell density and onset oscillation frequency at this maximal $\tau_s$ are given approximately by $N_o\simeq K(\alpha_1\alpha_2\epsilon)^{-1}$ and $\omega_o\simeq \epsilon\tau_y^{-1}$, respectively. This is compared to
$N_o\simeq K(\alpha_1\alpha_2)^{-1}$ and $\omega_o\simeq (\tau_a\tau_y)^{-1/2}$ at $\tau_s<(\tau_a\tau_y)^{1/2}$,
which are insensitive to $\epsilon$ as long as it is sufficiently small.
\\

\section{Glycolytic oscillations in yeast cells} 
\label{sect:glycolysis}

Glycolytic oscillations in dense yeast cell suspensions have been known for a long time~\cite{richard2003rhythm-S}. 
The phenomenon at cellular level is complex not only because of a large number of enzymes and metabolites involved, 
but also due to a multitude of regulatory interactions whose activation pattern and strength are not well understood.
Furthermore, flux diversion into side branches other than the main fermentative pathway can significantly 
attenuate or even diminish the oscillations. Yet the oscillations are easy to produce following the standard experimental protocols, 
suggesting that certain type of low dimensional mechanism inside a cell is at work.

\begin{figure}[!h]
\centering
\includegraphics[width=14cm]{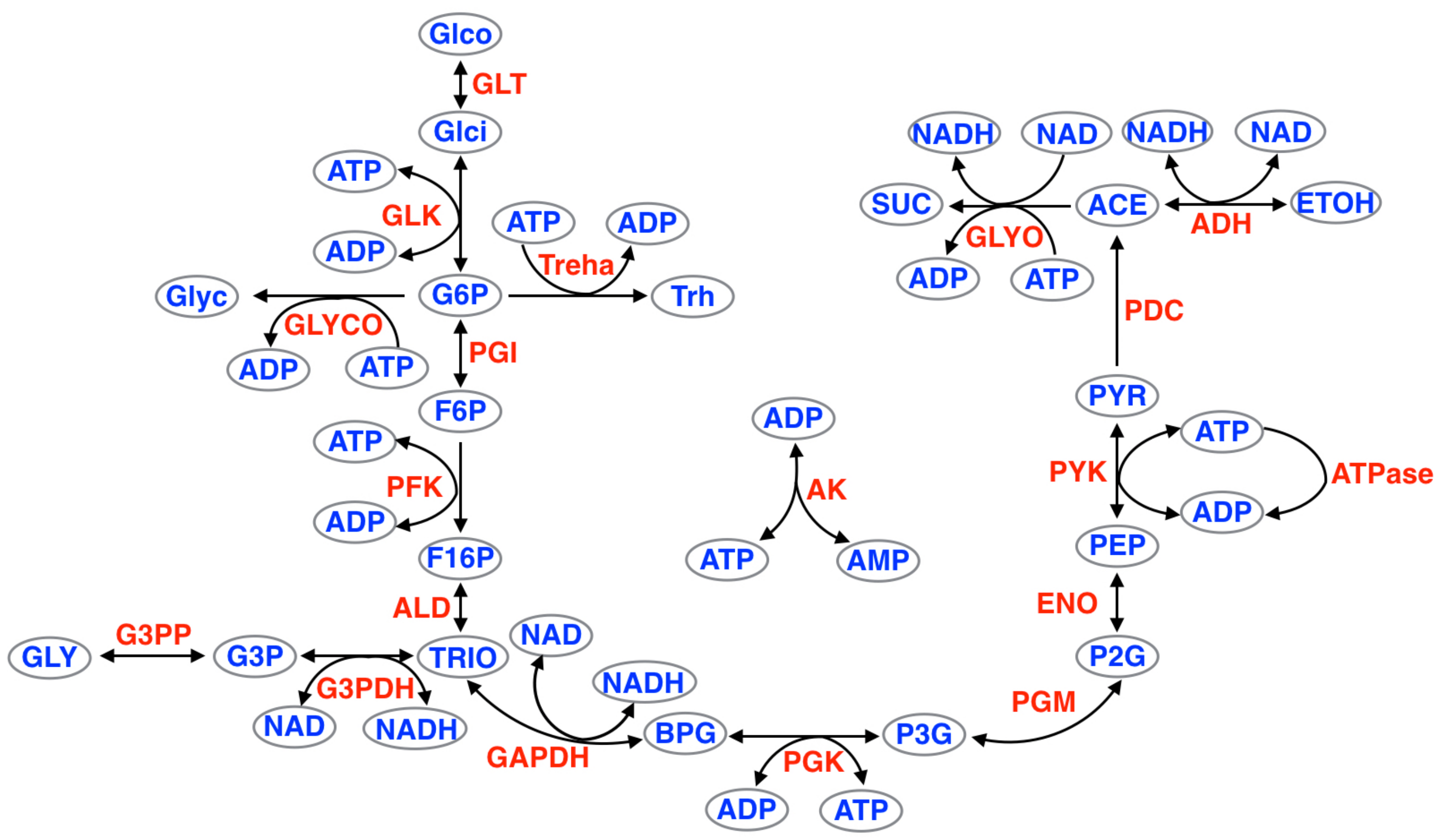}
\caption{\textbf{The network of reactions in a detailed model of glycolysis~\cite{du2012steady-S}}.  Letters in blue denote metabolites, 
while those in red are the reactions.  Directional (bidirectional) arrows indicate irreversible (reversible) reactions.   
Abbreviations:  Glco,  glucose;  ACE,  acetaldehyde,  ADH, alcohol dehydrogenase; AK, adenylate kinase; ALD, fructose-1,6-bisphosphate aldolase; BPG, 1,3-bis-phosphoglycerate; ENO, phosphopyruvate hydratase; F16P, fructose-1,6-bisphosphate; F6P, fructose 6-phosphate;  GAPDH, D-glyceraldehyde-3-phosphate dehydrogenase (phosphorylating); G3P, glycerol 3-phosphate; G3PDH, glycerol 3-phos- phate dehydrogenase; G6P, glucose 6-phosphate; GLYCO, glycogen branch; GLK, glucokinase (a hexokinase); P2G, 2-phosphoglycerate; P3G, 3-phosphoglycerate; PEP, phosphoenolpyruvate; PDC, pyruvate decarboxylase; PGI, glucose-6-phosphate isomerase; PFK, 6-phosphofructokinase; PGK, phosphoglycerate kinase; PGM, phosphoglycerate mutase; PYK, pyruvate kinase; PYR, pyruvate; Treha, trehalose branch; SUC, succinate branch; GLYO, glyoxylate shunt.}
\label{fig:kineticMap}
\end{figure}

In the following we explore the possibility of an adaptation route to yeast glycolytic oscillations. 
It is known that yeast cells communicate through the intercellular acetaldehyde (ACE) which acts as a redox signal. 
The intracellular redox ratio NAD/NADH affects the rate of the key reaction GAPDH separating ATP consuming and ATP
harvesting parts of the glycolytic pathway. Adaptation of the glycolytic flux to a rising (or receding) ACE level may result 
from  ACE's coupling to ATP homeostatic circuit on the time scale of seconds. 
We verify this scenario in a detailed model proposed by du Preez et al.~\cite{du2012steady-S} (referred to as the full model), 
and then develop a minimal model that helps us to understand the response phase diagram of the full model. 
At intermediate values of the extracellular ACE concentration, both the minimal model and the full model enter an oscillating
state. This part of the phase diagram is flanked by quiescent regions with adaptive response. The width of the adaptive
region can be tuned by altering side branches of glycolysis and downstream pathways, in particular inhibition of the glyoxylate shunt.
We then consider a system of coupled yeast cells, each metabolises according to the minimal model. The extracellular
ACE concentration is set by the cell volume fraction $\phi$. Collective oscillations at low cell densities result from synchronisation
of cells that oscillate on their own. At high cell densities, the elevated ACE level puts individual cells outside their oscillatory
regime when in isolation, yet the population as a whole may still oscillate collectively via the adaptation-driven DQS mechanism.
 The width of the oscillatory region can be significantly reduced on the low cell density side by cell-to-cell variations, 
and on the high cell density side by side reactions that reduce the adaptation accuracy of pyruvate pool which controls production of ACE. Our model study also highlights the importance of fast turnover of the extracellular ACE for sustained oscillations
as required by the phase-matching condition (Eq.~(4a), Main Text) and noted in previous experimental studies~\cite{richard1994yeast-S}.
\\

\begin{figure}[!h]
\centering
\includegraphics[width=15cm]{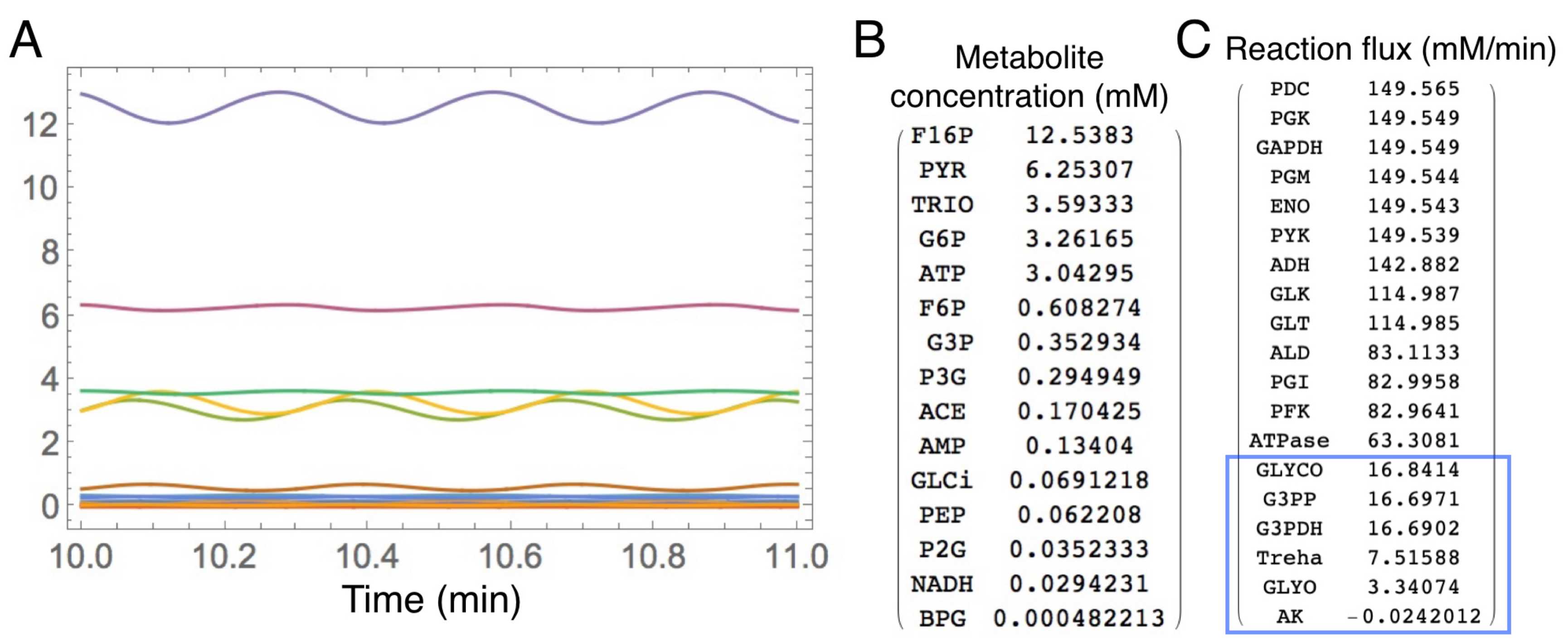}
\caption{\textbf{Spontaneous oscillations in the full model}.  
({\it A}) Trajectories of all  metabolites at glucose concentration Glco=10. 
({\it B}) and ({\it C}) Time-averaged metabolite concentrations and reaction fluxes in descending order.   }
\label{fig:unperturb}
\end{figure}

\subsection{Single-cell perturbation study}
 
The full intracellular reaction network of the kinetic model by du Preez et al.~\cite{du2012steady-S} is shown in 
Fig.~\ref{fig:kineticMap}. It contains around 20 reactions and 15 metabolite concentrations as dynamical variables. 
The reaction fluxes are highly nonlinear functions of these variables.   
Predictions of the model were shown to agree semi-quantitatively with experimental data on yeast glycolytic 
oscillations~\cite{du2012steady2-S}.  Below we use the same parameter values as adopted
in the original model termed dupree2 in~\cite{du2012steady-S}, unless otherwise stated.

\begin{figure}[!h]
\centering
\includegraphics[width=15cm]{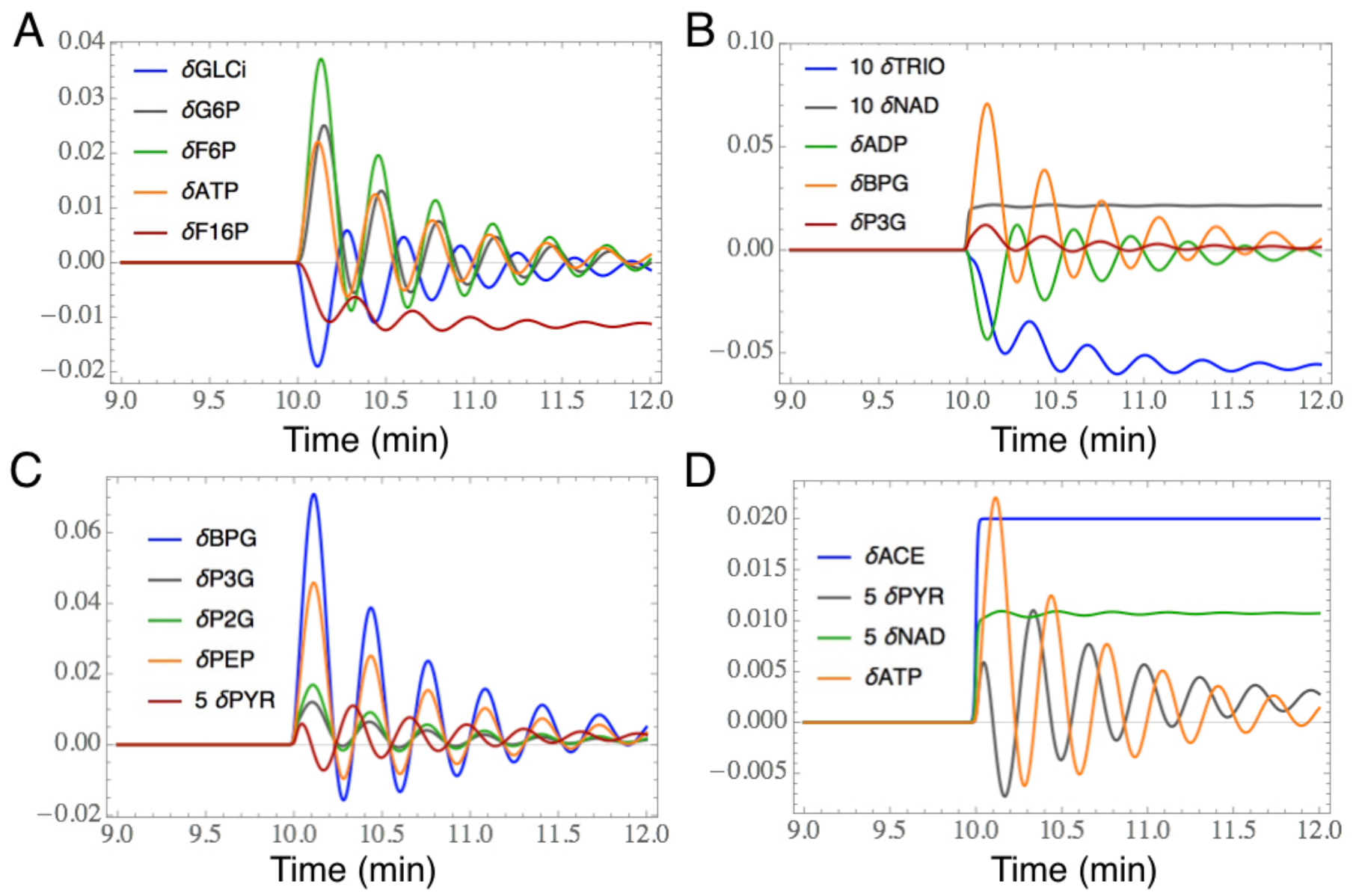}
\caption{\textbf{Response of metabolites to a redox signal at low ACE concentrations}. 
Here,  $\text{ACE}(t)=\text{ACE}_0[1+0.02 H(t)]$, with $H(t)$ being a Hill function with a large Hill coefficient.     
The notation $\delta x$ of a variable $x$ represents its relative change from a pre-stimulus level $\bar{x}$, i.e.,  
$\delta x\equiv [x(t)-\bar{x}]/\bar{x}$.  Quantities such as PYR and NAD which have too small values are amplified to make
them visible on the plot.   
({\it A}) The response of metabolites around G6P in the upper section of the glycolytic pathway;  
({\it B})  The response of metabolites around BPG in the middle section of the glycolytic pathway;  
({\it C}) The response of metabolites from BPG to PYR in the lower section of the glycolytic pathway;  
and ({\it D}) The response of metabolites in the downstream fermentation pathway.   
Parameters: $\text{ACE}_0=0.05$ and $\text{Glco}=10$.  }
\label{fig:adaptation-glycolysis}
\end{figure}


Fig.~\ref{fig:unperturb} shows an oscillatory solution of the model at the glucose concentration Glco =10 mM. 
The oscillation frequency is $\omega_0\approx 21$ min$^{-1}$, corresponding to a period of $\tau_0\simeq 0.3$ min. 
The mean concentration of ACE is 0.17 mM (Fig.~\ref{fig:unperturb}{\it B}). Reaction fluxes are concentrated along the linear
pathway from Glco to ETOH, while the side reactions carry much smaller flux  (Fig.~\ref{fig:unperturb}{\it C}, blue box).  
Below, we present response properties of the model using ACE concentration as the second control variable, in addition
to the extracellular glucose concentration. Time is measured in minutes and concentrations in mM.

To move out of the oscillatory regime, we lower the mean acetaldehyde concentration to $\text{ACE}_0=0.05$.
Experimentally, this can be achieved by adding cyanide (KCN) which reacts with ACE
in the solution~\cite{gustavsson2015entrainment-S}. 
Fig.~\ref{fig:adaptation-glycolysis} shows the time course of metabolites under a step-wise increase in the
ACE concentration. The four panels are organised following the order of metabolites along the glycolytic pathway, 
with the addition of ATP, ADP and NAD. 
Most metabolites adapt at least partially, except F16P and TRIO upstream of the reaction GAPDH
that uses NAD and NADH as cofactors. The redox pair NAD and NADH, being tightly connected to ACE, do not adapt either.

\begin{figure}[!h]
\centering
\includegraphics[width=15cm]{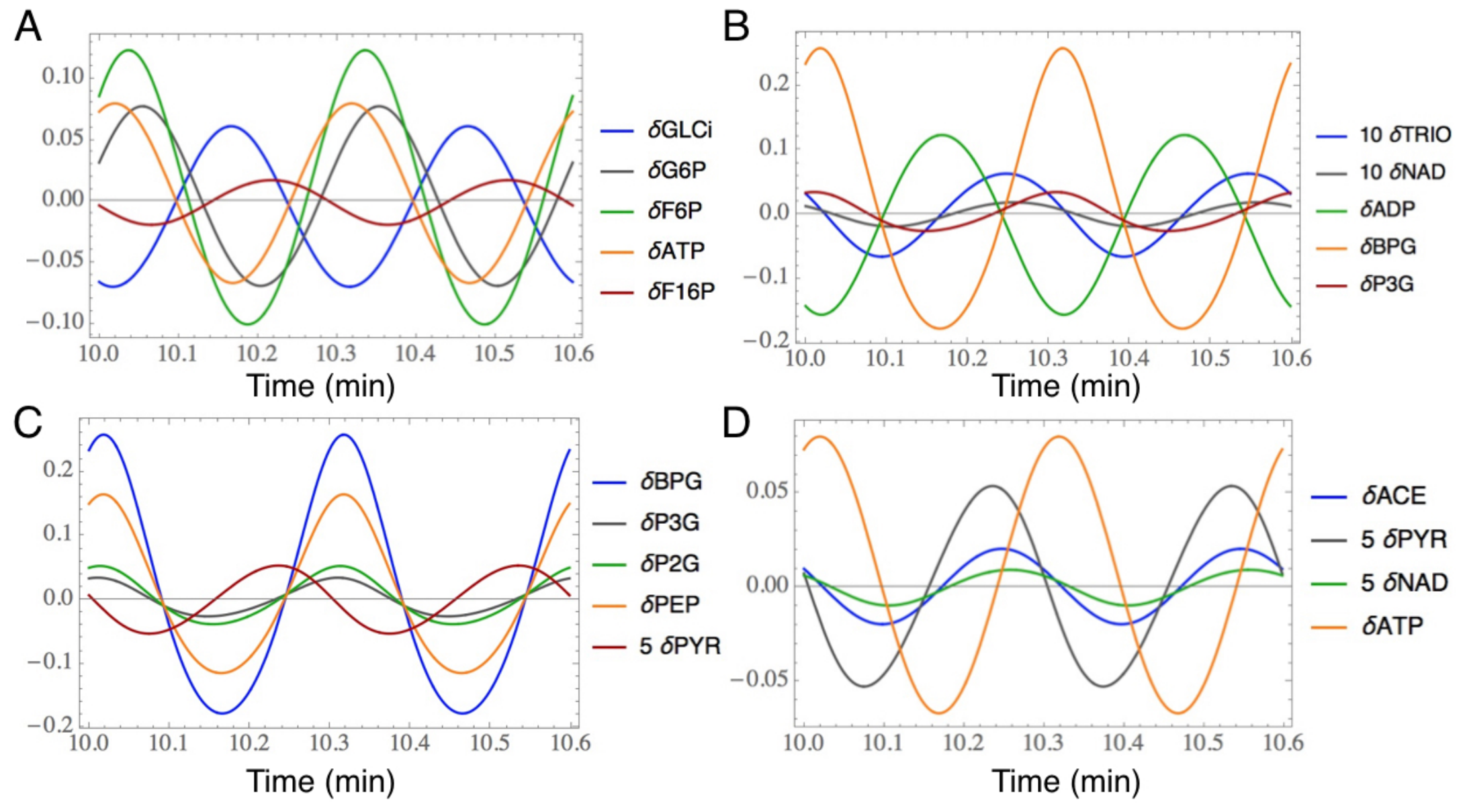}
\caption{\textbf{Concentration variations along the glycolytic pathway stimulated by a periodic redox signal}.  
Here, $\text{ACE}(t)=\text{ACE}_0[1+0.02 \sin(\omega t)]$. 
Organization of metabolites in panels ({\it A})-({\it D}) is the same as in Fig.~\ref{fig:adaptation-glycolysis}.
Parameters: $\text{ACE}_0=0.05$, $\text{Glco}=10$, and $\omega=21$.  }
\label{fig:periodic_perturbation}
\end{figure}

We now consider oscillations of the same set of metabolites stimulated by a periodic redox signal at the
frequency $\omega_0$ of spontaneous oscillations mentioned above. 
In Fig.~\ref{fig:periodic_perturbation}{\it A},  ATP, G6P and F6P are approximately in phase with each other,
but they are out of phase with Glci at the entry point of the pathway. The non-adaptive F16P has a behaviour of its own.   
The phase relations for these metabolites have been measured experimentally, and the results agree well with our numerics~\cite{richard1996sustained-S}.  
In Figs.~\ref{fig:periodic_perturbation}{\it B}-{\it C}, metabolites from BPG down to PEP share nearly the same phase
with each other and with ATP. The non-adaptive TRIO lags slightly behind
F16P. In Fig.~\ref{fig:periodic_perturbation}{\it D}, PYR at the end of the glycolytic pathway has an approximately 
$90^\circ$ phase lead over ATP, and furthermore a smaller phase lead over ACE and NAD.

\begin{figure}[!h]
\centering
\includegraphics[width=15cm]{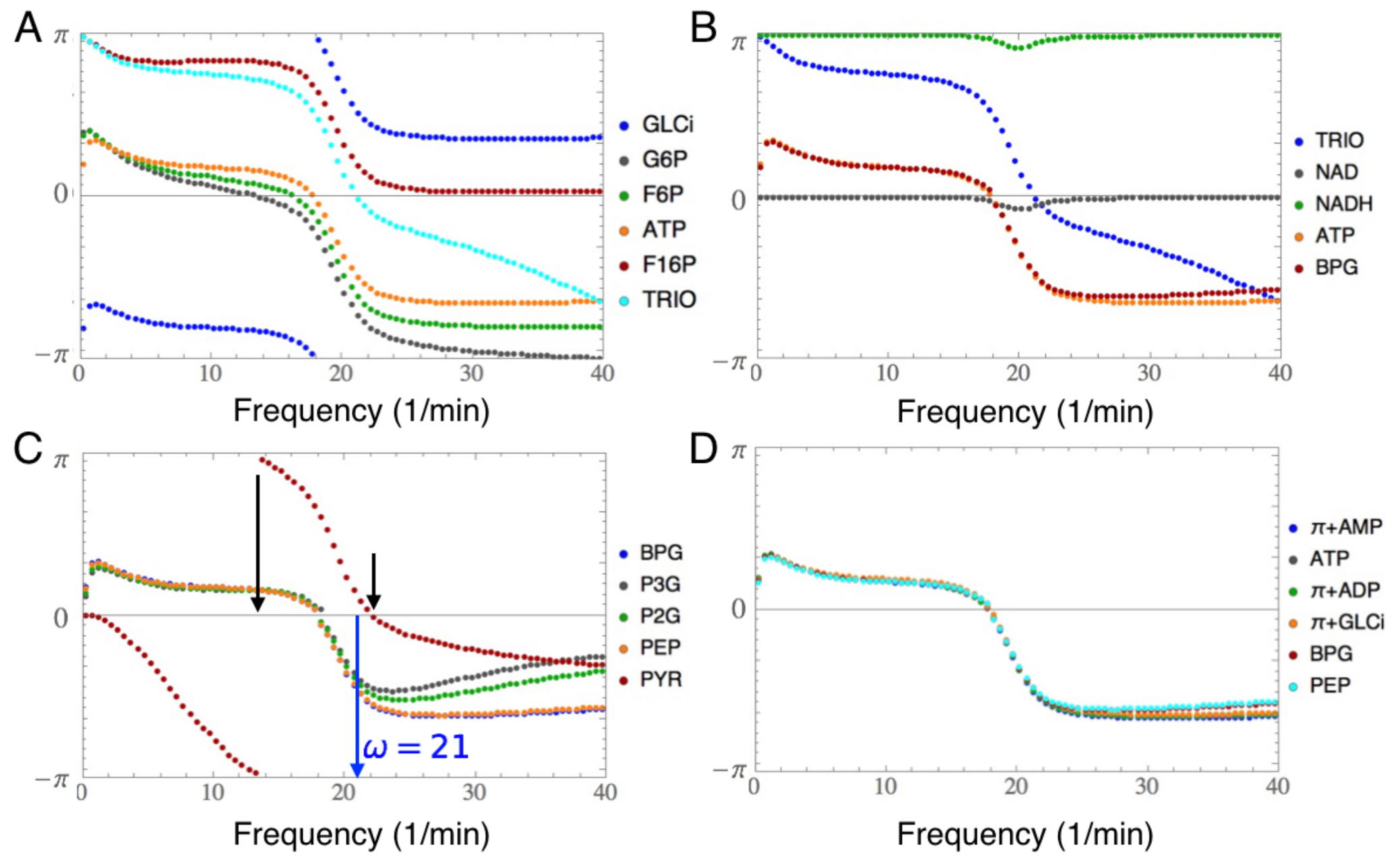}
\caption{\textbf{Phase shifts of metabolites against the frequency of a sinusoidal ACE signal}.  
({\it A}) Metabolites in the ``preparatory phase'' of the glycolytic pathway, where ATP is consumed to activate the 6-carbon ring
molecule. ({\it B}) Substrate, product and cofactors of the reaction GAPDH
that act as the receptor of the redox signal, together with ATP.
({\it C}) Metabolites in the ``payoff phase'' of the glycolytic pathway, where ATP is harvested. 
For the particular values of the extracellular glucose and acetaldehyde chosen,
phase lead of PYR over ACE occurs in the range of frequencies delimited by black arrows.
The blue arrow indicates the intrinsic frequency studied in Fig.~\ref{fig:periodic_perturbation}.  
({\it D}) Metabolites that appear in Eq.~(\ref{eq:phase-rule}). 
The phase shift of a number of metabolites shows a dip at the low frequency end, indicating a small but finite adaptation
error. Parameters: $\text{ACE}_0=0.05$, $\text{Glco}=10$. }
\label{fig:phase-relation}
\end{figure}

Fig.~\ref{fig:phase-relation} shows the phase shifts of metabolites against a sinusoidal signal ACE obtained
from our numerical simulations over a broad frequency range.   
Apart from PYR, the phase relationships among metabolites at $\omega_0$ hold also at lower frequencies. 
In Fig.~\ref{fig:phase-relation}{\it B}, it is seen that NAD has essentially the same phase as ACE in the frequency interval,
while NADH is completely out of phase. Therefore, on the timescale $\tau_0$, the phase information of ACE is passed 
without delay onto the redox ratio NAD/NADH, and fed into the network through the reaction GAPDH.   
Around $\omega_0$, the phase lead of NADH over ACE is slightly below 180$^\circ$,  
as observed in experiments on glycolytic oscillations~\cite{richard1996acetaldehyde-S}.    
Fig.~\ref{fig:phase-relation}{\it C} shows the downstream metabolites from BPG to PEP oscillate in phase with each
other for $\omega\leq\omega_0$, meaning the internal time scales for this part of the pathway are shorter than $\tau_0$.
In contrast, PYR develops a phase lead in the intermediate frequency regime,  as indicated by the two black arrows in Fig.~\ref{fig:phase-relation}{\it C}.  (Note that in Fig.~5{\it D} of the Main Text, the phase lead extends to
zero frequency indicating that the width of the regime depends on the glycolytic flux.)
The adaptive variable ATP also has a phase lead in the entire low frequency region.  

Fig.~\ref{fig:phase-relation}{\it D} shows the following phase relations between ATP and several other metabolites
as summarized by the equations below,
\begin{equation}
\label{eq:phase-rule}
 \phi_{ATP}=\pi+ \phi_{ADP}=\pi+ \phi_{AMP} \approx  \phi_{BPG}\approx  \phi_{PEP} \approx \pi+ \phi_{GLCi}.
\end{equation}
The first two relations among the nucleotides ATP, ADP and AMP simply reflect the conservation of their total number,
and that the fraction of AMP is much lower than the other two. 
In-phase relations apply to substrates BPG and PEP of the ATP harvesting reactions PGK and PYK, respectively,
while the out-of-phase relation is observed for GLCi in the ATP consuming reaction GLK. The fact that these relations
hold almost strictly in the entire frequency region suggests that quasi-steady-state conditions apply to these and neighbouring
reactions. It also suggests a prominent role of ATP in synchronising the phase of metabolites distributed along the glycolytic pathway.

In summary, our numerical results suggest the following mechanism of adaptation. Under a stepwise increase of ACE concentration,  
the information is passed with negligible delay to the redox ratio NAD/NADH, and then through the delayed reaction GAPDH 
to BPG and PEP, transiently boosting ATP production. The transient increase of ATP concentration then reduces the 
upstream glycolytic flux by inhibiting the reaction PFK, which in turn decreases the downstream TRIO concentration,
eventually returning the GAPDH flux to its pre-stimulus level.  
Although many metabolites adapt, the negative feedback loop of ATP production appears to be the core.
Fig.~5{\it B} in the Main Text shows a more complete phase diagram of the response properties at other values of 
$\text{ACE}_0$ and $\text{Glco}$ concentrations, including the region of spontaneous oscillations.

\subsection{A minimal model for glycolytic oscillations}

\begin{figure}[!h]
\centering
\includegraphics[width=13cm]{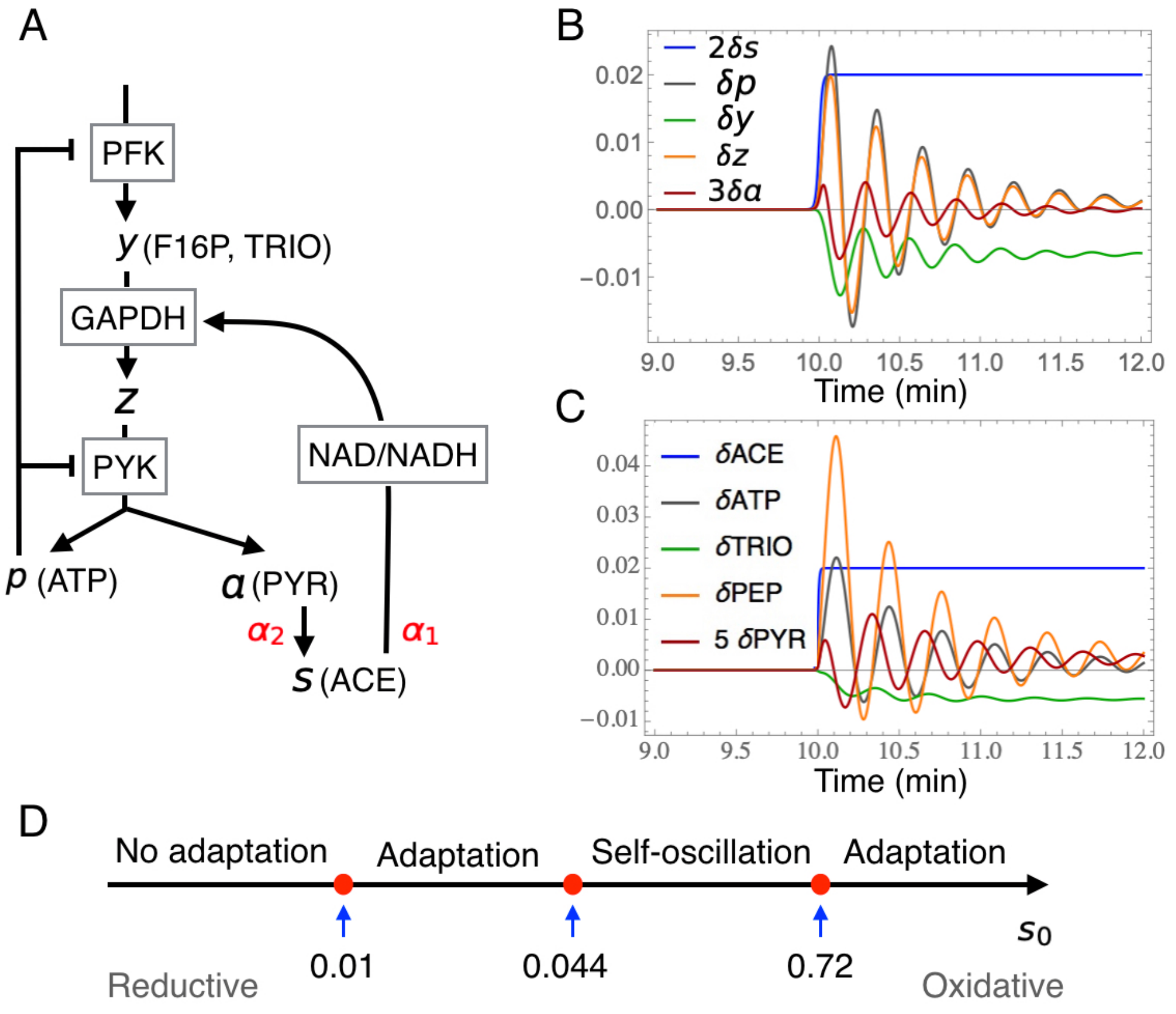}
\caption{\textbf{A minimal model of glycolysis with redox control}.  ({\it A}) The network of metabolites (symbols)
and reactions (boxes).  ({\it B}) Response of metabolites upon a stepwise perturbation $s(t)=0.04 (1+0.01 H(t))$.  
({\it C}) Response of corresponding metabolites in the full model computed using parameter values given in 
FIG.~\ref{fig:adaptation-glycolysis}.   
({\it D}) Response properties of the minimal model as the intracellular redox state changes from reductive to oxidative
(left to right). Parameters: $h=3, \alpha_2=\alpha_1=1$,  $\tau=0.01$, $c_0=0.02$, and $\epsilon=0.01$.  }
\label{fig:MinimalModel}
\end{figure}

We constructed a minimal model to test various quantitative aspects of the adaptation mechanism described above.
Reduction in the number of dynamic variables is achieved by lumping consecutive metabolites along the linear pathway 
that are phase synchronised into a single variable denoting their total concentration. 
This is a reasonable approximation when interconversion among these metabolites is much faster than 
the time of interest, e.g. the oscillation period.
Fig.~\ref{fig:MinimalModel}{\it A} illustrates the selected variables and their interactions.  
Here, $y$ represents intermediate metabolites that do not adapt (F16P and TRIO), thereby playing the role of a memory node. 
The variable $z$ represents metabolites from BPG to PEP along the glycolytic pathway. 
The ATP concentration is denoted by $p$, while the concentration of PYR,
substrate for the ACE producing reaction PDC and thus the corresponding cell activity here, is denoted by $a$. 
Since NAD (NADH) is always in phase (out of phase) with ACE, we adopt the redox ratio NAD/NADH 
as the signal $s$ instead. Motivated by a phenomenological two-component model for glycolytic oscillations  
in Ref.~\cite{chandra2011glycolytic-S}, we introduce a minimal model of glycolysis with redox control as follows:
\begin{subequations}\label{eq:MinimalModel}
\begin{eqnarray}
\tau\dot{y}&=&\frac{2p}{1+p^{2h}}-(\alpha_2 s+c_0) y -\epsilon y  ,\\
\tau \dot{z}&=& (\alpha_2 s+c_0) y - \frac{2z}{1+p^2},\\
\tau \dot{p} &=&-\frac{2p}{1+p^{2h}}+2  \frac{2z}{1+p^{2}}-\frac{2p^2}{1+p^2}.
\end{eqnarray}
Here, $2 p/(1+p^{2h})$ gives the reaction flux of PFK that consumes ATP and is also inhibited by ATP at high concentrations
(i.e., the negative feedback loop), with the inhibition strength set by the exponent $h (>1/2)$. The entry carbon flux into
the glycolysis pathway is assumed not to be rate limiting, e.g.., one is in a situation of high extracellular glucose concentration.
The term  $(\alpha_2 s+c_0) y$ gives the reaction flux of GAPDH, where  $c_0$ sets the ``basal'' enzyme velocity at $s=0$.  Leakage of TRIO into the side branch is represented by $\epsilon y$.
The term $2z/(1+p^2)$ gives the reaction flux of PYK (and also PGK), which produces ATP but is also inhibited by ATP.   
In Eq.~(\ref{eq:MinimalModel}c), the stoichiometric factors 1 and 2 in the first two terms
on the right-hand-side correspond to the ATP consumption and production upstream and downstream of
TRIO, respectively.  ATP consumption by the cell outside of glycolysis (e.g., ATPase activity) is modelled by the term 
$2p^2/(1+p^2)$, which grows with the ATP concentration until saturation at a maximal value 2.  
The output variable $a$ (PYR) is produced by the same flux that produces $p$ (ATP) and degraded at a constant
rate $\alpha_1$,
\begin{equation}
\tau \dot{a}=\frac{2z}{1+p^2}-\alpha_1 a.
\end{equation}
\end{subequations}
The glyoxylate shunt GLYO, which is active at ACE$_0\simeq 0.2$ mM or above in the full model, is turned off.
For simplicity, we have chosen the time constants on the left-hand-side of the equations to be the same. As we show below,
this choice is adequate for recovering the main low frequency properties of the full model.

Fig.~\ref{fig:MinimalModel}{\it B} shows the response of the dynamical variables to a sinusoidal redox variation
centred around $s_0=0.04$. Except the buffer variable $y$, all other variables show adaptive behaviour, with
$a$ gaining a phase lead of $90^\circ$ over $p$.  
For comparison, we show in Fig.~\ref{fig:MinimalModel}{\it C} the response properties of corresponding metabolites
in the full model in the adaptive regime, which are indeed quite similar.
We have examined the response properties of the minimal model at other values of $s_0$
and identified four qualitatively different regimes as shown in Fig.~\ref{fig:MinimalModel}{\it D}.
As in the case of the full model with sufficient glucose (Main Text, Fig. 5{\it B}), spontaneous oscillations (i.e., limit
cycle solution) occur at intermediate values of $s_0$, flanked by adaptive but non-oscillatory regions.

We note in passing that the two-component model of Chandra {\it et al.}~\cite{chandra2011glycolytic-S}
also exhibits spontaneous oscillations when the rate constant $k$ of the pyruvate kinase reaction 
(PYK in Fig.~\ref{fig:MinimalModel}{\it A}) takes on intermediate values. As $k$ affects the delay
time of the negative feedback control in ATP production, in this sense it plays a similar role as $s_0$.
However, our model contains an additional buffer node TRIO which is necessary for the adaptive behaviour seen in
Fig.~\ref{fig:MinimalModel}. We have also made the ATP consumption rate dependent on the ATP concentration
to eliminate certain pathological aspects of the Chandra {\it et al}. model at low values of $p$.
Furthermore, our numerical analysis suggests that a sufficiently small but finite adaptation error $\epsilon$ associated
with low flux diversion is needed to reproduce the response diagram Fig.~\ref{fig:MinimalModel}{\it D}.
On the high (oxidative) end of $s_0$, the reaction GAPDH drives down $y$ (TRIO) and hence the flux of the side reaction,
making the system adaptive even when $\epsilon\sim 1$.

\begin{figure}[!h]
\centering
\includegraphics[width=14cm]{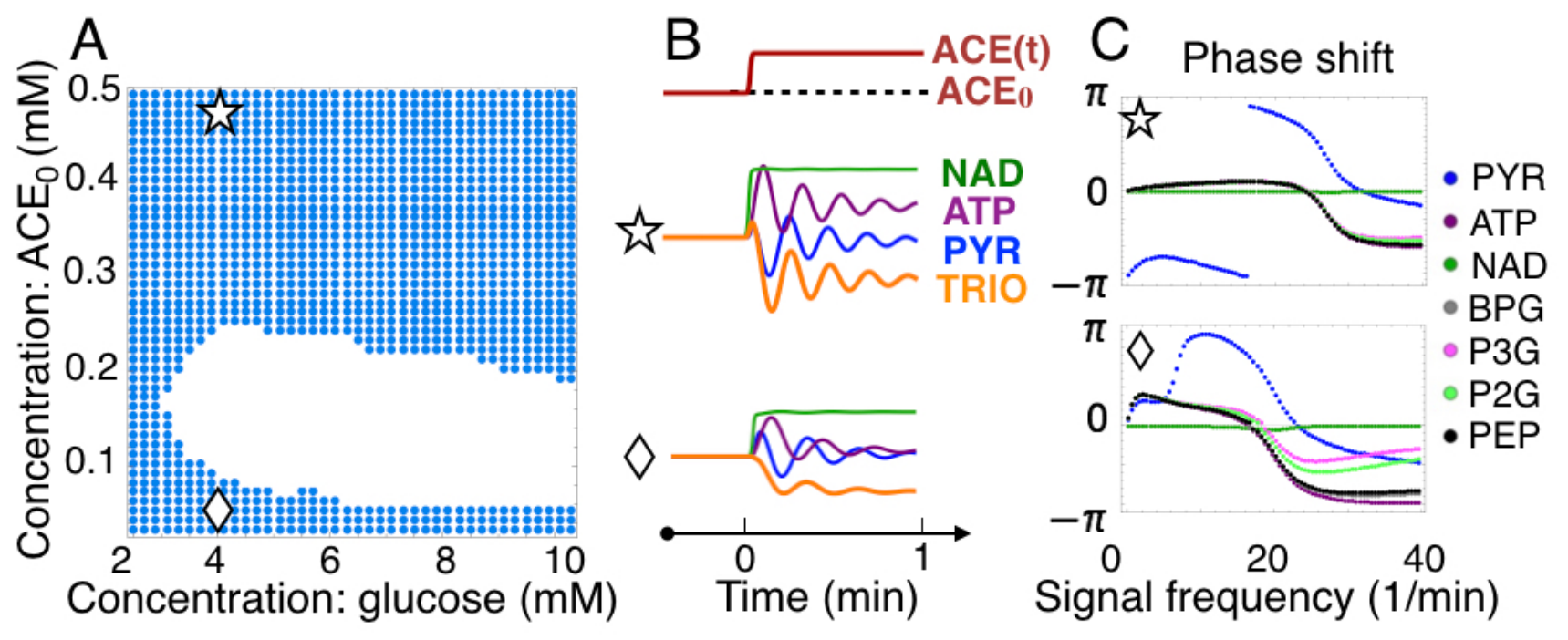}
\caption{\textbf{Phase diagram of the modified glycolysis model with the GLYO reaction switched off}.  
Now PYR and ATP both adapt over a broad region of the phase diagram, in agreement with
the minimal model. }
\label{fig:phase_diagram_glycolysis_removing_SUC}
\end{figure}

Comparing the response diagrams of the minimal model (Fig.~\ref{fig:MinimalModel}{\it D}) and 
of the full model at high extracellular glucose concentrations (Main Text, Fig. 5{\it B}), we see that the adaptive regime 
on the oxidative side is restricted to a much narrower region in the latter case. 
Upon a detailed investigation of the full model we found that, at higher values of 
ACE$_0$, the side reaction GLYO is activated. Shutting down the reaction, we obtained a response diagram similar to that of the minimal model (Fig.~\ref{fig:phase_diagram_glycolysis_removing_SUC}). 
The reaction GLYO uses NAD as cofactor and consumes ATP (see Fig.~\ref{fig:kineticMap}).
With regard to the change in NAD/NADH ratio upon an upshift of ACE, it has an opposite effect as compared to ADH.
This and ATP consumption by GLYO leads to a sign reversal in the transient response of ATP to ACE upshift 
at ACE$_0\simeq 0.2$ mM in the full model (Fig. 5{\it B} in the Main Text).
Inhibition of GLYO eliminates the sign switch and makes PYR adapting to ACE over a much larger region of the phase diagram
(Fig.~\ref{fig:phase_diagram_glycolysis_removing_SUC}).


\begin{figure}[!h]
\centering
\includegraphics[width=16cm]{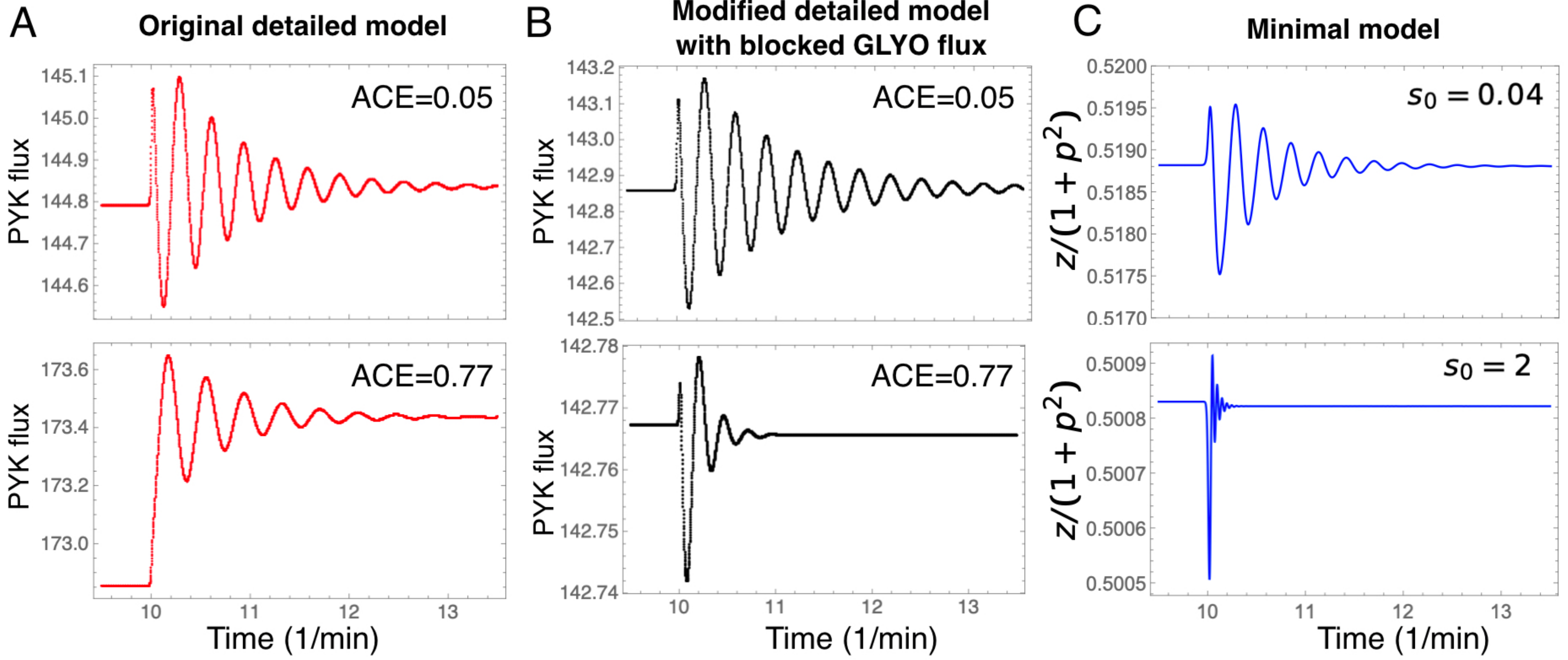}
\caption{\textbf{The response of PYK flux to a step perturbation of ACE at $t=10$}.  
({\it A}) The response of the original full model at ACE=0.05 and 0.77, respectively.  
({\it B}) The response of the modified full model with blocked GLYO reaction at ACE=0.05 and 0.77, respectively. 
({\it C}) The response of the minimal model.  
Parameters: Glco=10 for ({\it A}) and ({\it B}); the parameters for the minimal model are the same as in Fig.~\ref{fig:MinimalModel}.  }
\label{fig:glycolysis_PYK_flux}
\end{figure}

Fig.~\ref{fig:glycolysis_PYK_flux} shows representative time courses of the PYK reaction flux to a stepwise ACE signal, 
computed using the original and modified glycolysis model, as well as the minimal model. 
Concentration of its product, PYR, is found to be proportional to the PYK reaction flux in all three models, i.e., 
the degradation rate of PYR is a constant.  
The original and modified models exhibit nearly identical adaptive response on the low ACE (reductive) side, 
but differ on the high ACE (oxidative) side. In the latter case, the PYK flux is significantly higher and also non-adaptive
when the glyoxylate shunt (GLYO) is on. Further numerical investigations of the 
full model with blocked GLYO reaction show that it shares the following features of the minimal model
as the oxidation level increases: 1) the frequency inside the oscillatory regime increases; 
2) (mean) $p$ (ATP) and $z$ (BPG, P3G, P2G, and PEP concentrations) increase by a moderate amount; 
3) $y$ (TRIO and F16P concentrations) decreases; 4) $a$ (PYR concentration) first increases, then decreases.  
Experimental time-course measurement with blocked glyoxylate shunt will serve to validate or improve the model assumptions.

\subsection{Glycolytic oscillation in coupled yeast cells}
 
To study collective oscillations in a population of cells whose internal dynamics follows Eqs. (\ref{eq:MinimalModel}),
we adopt the following signal dynamics as in Ref. \cite{wolf2000transduction-S}: 
\begin{subequations}
\begin{eqnarray}
\label{eq:signal-glycolysis-a}
\tau_s \dot{s}_{in}&=&\alpha_1 a-k_{in} s_{in}- D (s_{in}-s_{ex}), \\
\label{eq:signal-glycolysis-b}
\tau_s \dot{s}_{ex} &=& \phi D(s_{in}-s_{ex})-k_{ex} s_{ex}. 
\end{eqnarray} 
\label{eq:signal-glycolysis}
\end{subequations}
Here $s_{in}$ and $s_{ex} $ are the intracellular and extracellular signal concentration, respectively; $D$ is the membrane 
permeability of the signalling molecule; $k_{in}$ and $k_{ex}$ are the intracellular and extracellular signal degradation rate; 
and $\phi$ is the volume fraction of yeast cells, which increases with the cell density, and saturates at 1. The extracellular
signal strength (i.e., acetaldehyde concentration) in the coupled system is a function of $\phi$.

Let us first consider the situation of fast equilibrium between $s_{in}$ and $s_{out}$. Previously, Silvia De Monte \emph{et al.} 
proposed a diffusion timescale $\tau_s\approx 0.003$ s by assuming a quasi-stationary concentration profile and that 
ACE molecules need to diffuse across a spherical shell with an inner radius $r_1=3\ \mu$m and an outer radius 
$r_2=6.5\ \mu$m~\cite{de2007dynamical-S}. This diffusion timescale is much smaller than the oscillation period of $37$ s.  
Assuming the time for an ACE molecule to cross the cell membrane is of the order of 1 s or less,
we obtain the following approximate equation for $s=(s_{in}+s_{ex})/2$,  
\begin{equation}
\tau_s \dot{s}=\frac{\phi}{1+\phi} \alpha_1 a-\Big(\frac{\phi k_{in}+k_{ex}}{\phi+1 }\Big)s.
\label{eq:reduced-signal-dynamics}
\end{equation} 
Up to corrections of order $\epsilon$, the stationary state of the dynamical system defined by Eqs.~(\ref{eq:reduced-signal-dynamics})
and (\ref{eq:MinimalModel}) is given approximately by, 
\begin{equation}
p\approx 1, \quad z\approx 1,\quad y\approx \frac{1}{ \alpha_2s + c_0},\quad  a\approx \frac{1}{ \alpha_1},\quad s\approx \frac{\phi }{ \phi k_{in}+k_{ex} }.
\end{equation}
The signal strength increases with the volume fraction and saturates at $1/(k_{in}+k_{ex})$.  
At small but finite $\epsilon$, corrections to the above expressions become significant at large $y$ or small $s$,
where the side reaction G3PDH in Fig.~\ref{fig:kineticMap} is activated to divert the glycolytic flux.
In the numerical studies presented below, we set the two ACE degradation rates $k_{in}$ and $k_{ex}$ to be small.
The signal strength $s$ varies over a broad range as the cell volume fraction $\phi$ increases. 

\begin{figure}[!h]
\centering
\includegraphics[width=14cm]{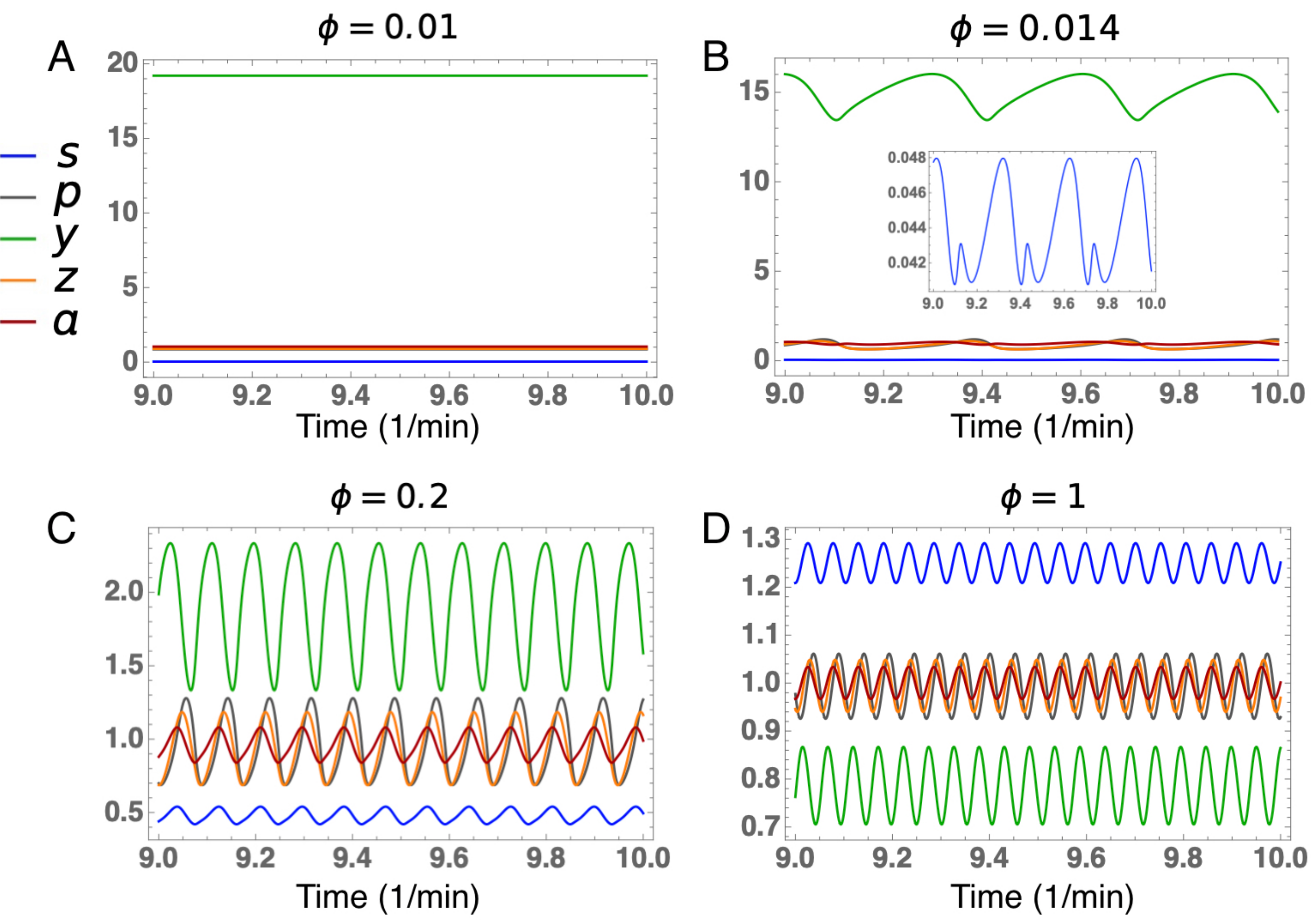}
\caption{\textbf{Collective dynamics of the minimal model coupled via Eq.~(\ref{eq:reduced-signal-dynamics})}.  
({\it A})-({\it D}) Temporal trajectories at selected values of the volume fraction $\phi$.  The same color scheme of 
variables is used. Inset in {\it B} shows the signal trajectory on an enlarged scale.  
Parameters:  $h=3,\alpha_2=\alpha_1=1, \epsilon=0.01$,    
$k_{in}=0.5$, $k_{ex}=0.3$, $\tau=0.01$, $c_0=0.02$, and $\tau_s=0.001$. }
\label{fig:collective_oscillation_glycolysis_illustration}
\end{figure}

Fig.~\ref{fig:collective_oscillation_glycolysis_illustration} shows numerical solutions of the coupled minimal model at 
four selected $\phi$ values. Except the case at $\phi=0.01$, oscillations of $s$ and the intracellular variables are seen.
In Fig.~\ref{fig:collective_oscillation_glycolysis_phase_diagram}, we plot the 
oscillation amplitudes and time-averaged values of $s$ and $O$ against the cell volume fraction $\phi$.
From the lower panel of Fig.~\ref{fig:collective_oscillation_glycolysis_phase_diagram}{{\it A} we see that, 
for the signal dynamics chosen, the lower adaptive regime in Fig.~\ref{fig:MinimalModel}{\it D} is mapped to a
narrow interval of cell volume fraction $0.003<\phi<0.013$. 
From Fig.~\ref{fig:collective_oscillation_glycolysis_illustration}, we see that onset of collective oscillations in the 
coupled system takes place somewhere between $\phi=0.01$ and $0.014$. More detailed studies indicate that the transition
is not the expected Hopf bifurcation type, but instead emergence of a limit cycle at finite amplitude.
Similar behaviour was seen in the study of the full kinetic model (see Fig. 10 in~\cite{du2012steady2-S}).  
On the other hand, experimental work seem to support the Hopf bifurcation scenario~\cite{dano1999sustained-S,de2007dynamical-S}. 
We leave this issue to future investigations.

\begin{figure}[!h]
\centering
\includegraphics[width=14cm]{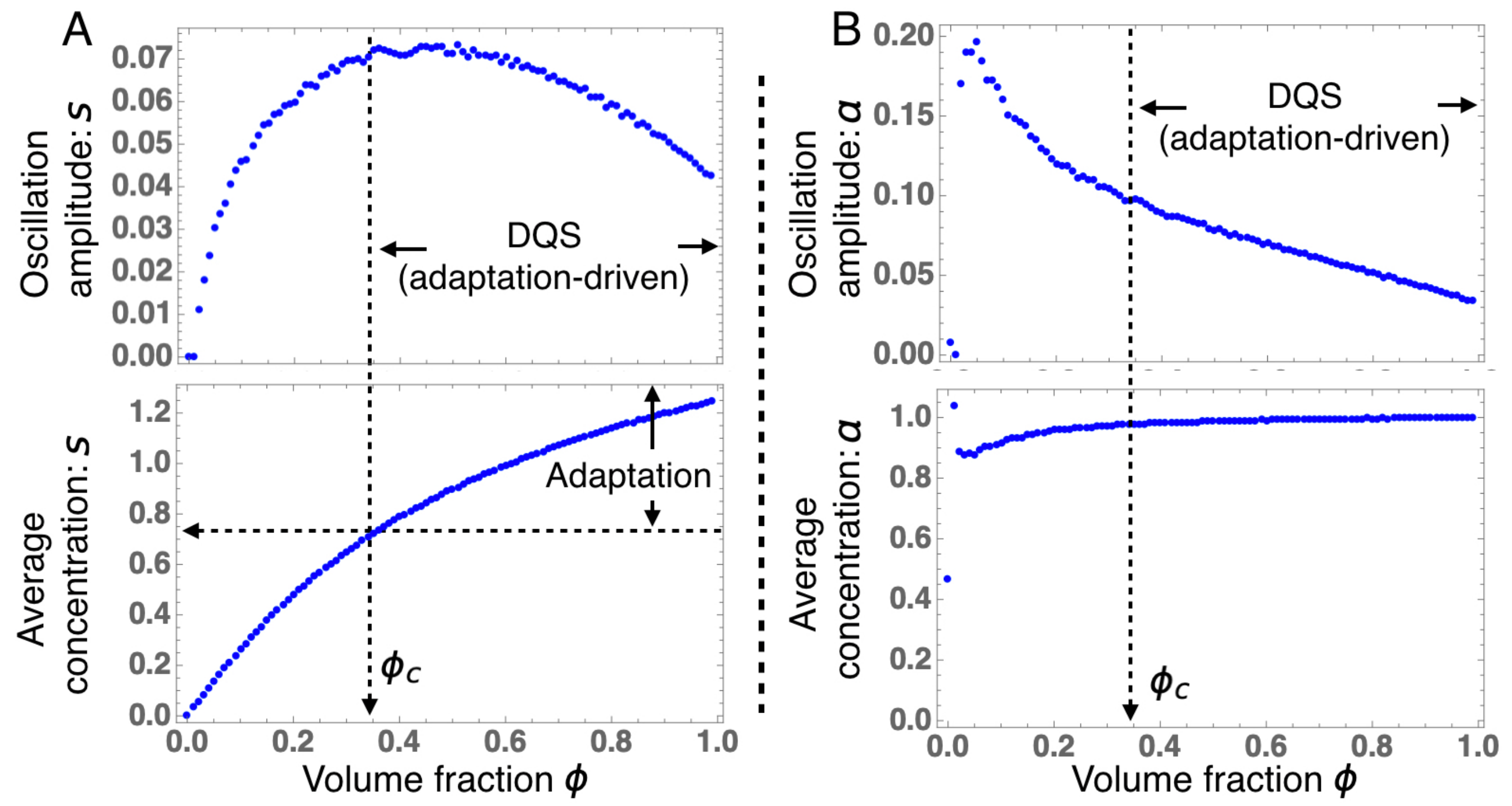}
\caption{\textbf{Collective oscillations against the yeast cell density}.  
({\it A}) Upper panel: oscillation amplitude of the signal $s$ as a function of the cell volume fraction $\phi$.  
Lower panel: time-averaged signal concentration against $\phi$.  At $\phi_c=0.34$, the signal strength reaches the upper 
threshold $s_c=0.72$ for the oscillating state of individual cells (Fig.~\ref{fig:MinimalModel}{\it E}).
The oscillating state at $\phi>\phi_c$ can be considered as DQS driven by adaptation. 
({\it B})  Upper panel: oscillation amplitude of the sender node $a$ against $\phi$.  
Lower panel: time-averaged value of $a$ against $\phi$.   Parameters are the same as in Fig.~\ref{fig:collective_oscillation_glycolysis_illustration}.}
\label{fig:collective_oscillation_glycolysis_phase_diagram}
\end{figure}

Beyond the onset point, oscillation amplitudes vary continuously with the cell density.
For $\phi>\phi_c=0.34$, the time-averaged value of $s$ falls in the upper adaptive regime in Fig.~\ref{fig:MinimalModel}{\it E}.
Since the cell density here already exceeds the threshold value required for collective behaviour of adaptive units,
oscillations continue.

Finally, we present numerical results demonstrating the effect of a slower cross-membrane transport of acetaldehyde 
on the collective dynamics. The system dynamics is defined by Eqs.~(\ref{eq:MinimalModel})] for individual cells (with $s=s_{in}$)
together with Eqs.~(\ref{eq:signal-glycolysis}) for the intracellular and extracellular signal concentrations.  
Fig.~\ref{fig:glycolysis_glycolysis_effect_D} shows the oscillation amplitude of $s_{in}$ together with the time-averaged
values of $s_{in}$ and $s_{out}$ at selected values of $D$. At $D=100$ and $10$, $s_{in}$ and $s_{out}$ are nearly identical
and the system behaviour is essentially the same as described above under the fast equilibrium assumption.
At $D=1$, the time-averaged value of $s_{ext}$ is noticeably smaller than that of $s_{in}$, indicating a significant
gradient of acetaldehyde concentration across the cell membrane. Nevertheless, collective oscillations via DQS continue over a broad range of the cell density.  Interestingly, the oscillation is arrested at very high densities. 
Collective oscillations disappear at $D=0.1$. Here, $s_{in}$ remains high due to the slow
intracellular degradation rate $k_{in}$, which places the single-cell dynamics in the upper adaptive regime even when the
cell density is very low. However, the phase delay across the cell membrane changes the response properties of the cell
to external signal variations. In this case, $s_{in}$ should be considered as the sender of the external signal but as one
can see from Eq. (\ref{eq:signal-glycolysis-a}), the adaptation
of $a$ to $s_{in}$ does not translate to adaptation of $s_{in}$ to $s_{ex}$ when $D$ is small. The latter is required for the adaptation route to collective oscillations.
 
\begin{figure}[!h]
\centering
\includegraphics[width=16cm]{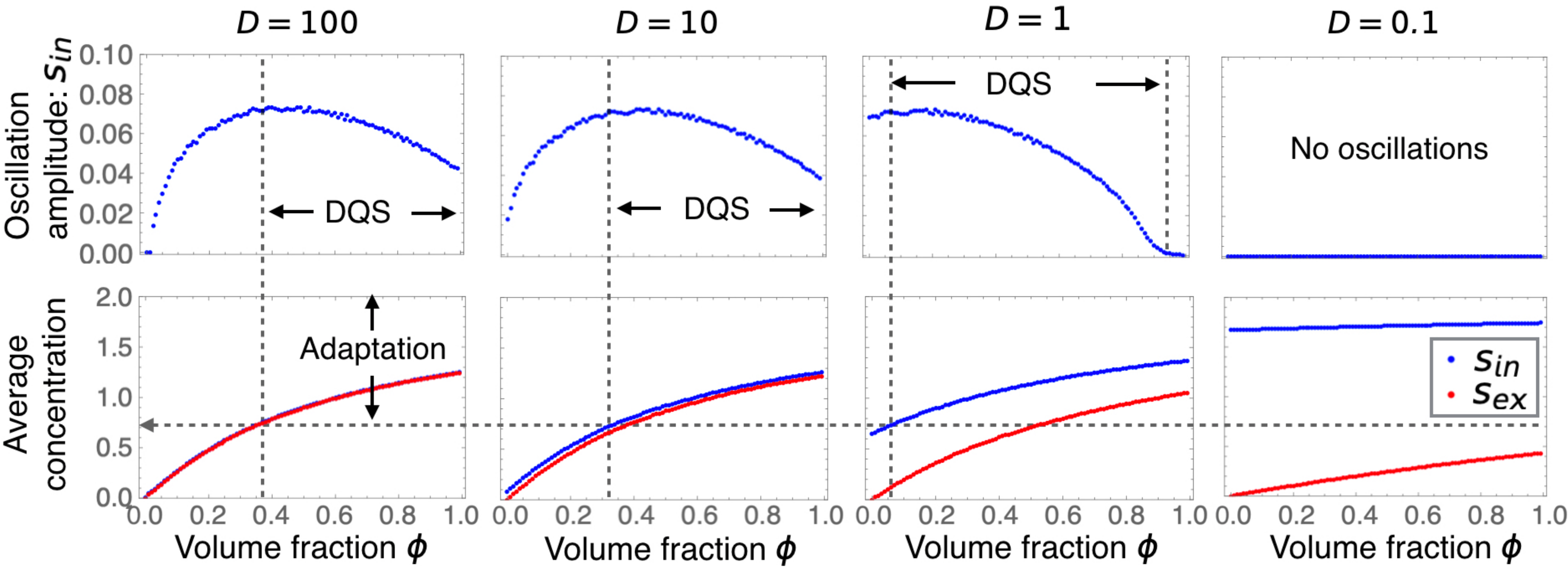}
\caption{\textbf{Effect of delay in cross-membrane transport of the signalling molecule on collective dynamics}.  
Results of numerical integration of Eqs.~(\ref{eq:MinimalModel})] coupled to the two-component signal dynamics 
Eq.~(\ref{eq:signal-glycolysis})] at selected values of $D=100,10,1, 0.1$.   
In the lower panels, the blue and orange dots correspond to the average of $s_{in}$ and $s_{ex}$, respectively.  
Other parameters are the same as in Fig.~\ref{fig:collective_oscillation_glycolysis_illustration}.}
\label{fig:glycolysis_glycolysis_effect_D}
\end{figure}

In summary, under fast equilibration between intracellular and extracellular acetaldehyde concentrations, the coupled system 
exhibits collective oscillations over a broad range of cell densities, encompassing the adaptive and oscillatory regimes of a single cell.
Onset of collective oscillations at low cell densities exhibit complex behaviour due to the assumed sensitivity of the
reaction GAPDH to the NAD/NADH ratio. Delay in the cross-membrane transport of acetaldehyde weakens
adaptation of intracellular metabolite concentrations to change in the extracellular acetaldehyde concentration,
and may eliminate collective oscillations altogether when the delay is too long~\cite{richard1994yeast-S}.
At moderate delays, rise in the intracellular
acetaldehyde concentration brings individual cells to the oscillatory even when in isolation. The enhanced oscillation amplitude
at $D=1$ and low cell densities seen in Fig.~\ref{fig:glycolysis_glycolysis_effect_D}, however, is obtained under the assumption
that all cells in the population behave identically. This behaviour is susceptible to cell-to-cell variations as well as temporal noise
in intracellular dynamics. Our model study exposes this and other subtleties that can affect emergence of collective oscillations.
The specific effects we identified in this work could serve to guide the design of future experiments where various model
parameters can be controlled quantitatively, e.g., $k_{ex}$ for extracellular degradation rate of acetaldehyde by adjusting
the flow rate in microfluidic setups~\cite{gustavsson2015entrainment-S}.


\end{document}